\documentclass[reprint, aps, prl]{revtex4-2}
\usepackage{style/my_packages}

\begin{document}


\title{Tailored Error Mitigation for Single-Qubit Magnetometry}

\author{Miriam Resch}
\altaffiliation{These authors contributed equally.}
\author{Dennis Herb}
\altaffiliation{These authors contributed equally.}
\author{Mirko Rossini}
\author{Joachim Ankerhold}
\author{Dominik Maile}

\affiliation{Institute for Complex Quantum Systems, Ulm University, 89069 Ulm, Germany}
\affiliation{Center for Integrated Quantum Science and Technology (IQST) Ulm-Stuttgart, Germany}

\begin{abstract}
Quantum sensing is an emerging field with the potential to outperform classical methods in both precision and spatial resolution. 
However, the sensitivity of the underlying quantum platform also makes the sensors highly susceptible to their environmental noise.
To address this issue, techniques from the
field of quantum error mitigation use information about the noise to improve measurement results.
We present a novel mitigation technique for quantum sensors to efficiently reverse the effects of any noise that can be described by a completely positive trace preserving map.
The method leverages the knowledge acquired by a pre-characterization step of the device to automatically adapt to the complexity of the dissipative evolution and
to indicate optimal sensing times $\tau$ to achieve the most accurate results. We demonstrate that our method reaches the best achievable sensitivity in noisy single-NV-center magnetometry. 
 This work marks a further step toward more resilient quantum sensors with the smallest scale of resolution.

\end{abstract}

\maketitle


\paragraph{Introduction--}
Quantum sensing is one of the most promising applications of quantum technologies, alongside quantum computing and communication. Its possible applications range from condensed matter physics to neuroscience, and biological systems \cite{schirhagl_nitrogen-vacancy_2014, Degen2017,barry_sensitivity_2020, Aslam2023}.
Despite much effort being focused on reaching the lowest  sensitivities by engineering sensors made of multiple (entangled) quantum units (viz. photons, qubits, or spins), improving spatial resolution is equally important and is achieved by advancing single-qubit sensing devices. These single-qubit sensors—such as single nitrogen–vacancy (NV) centers in diamond \cite{WrachtrupsingleNV,Balasubramanian2008,Maletinsky2012} —are essential for the development of novel
sensing technologies capable of measuring molecular-scale, room-temperature systems, e.g. monitoring a cell’s metabolism in vivo \cite{Aslam2023}. \\
In practice, the sensitivity of a quantum sensing device is systematically hindered by the interaction with the surrounding environment, causing decoherence, relaxation and therefore loss of information
often characterized by multiple time constants 
\cite{Plenio2012}.
Especially single NV-centers in practical settings  experience complex noise interferences from interactions with local impurities, such as spins from surface electrons or nearby nuclei.
The resulting dissipative evolution is described by non-trivial quantum channels, leading to multi-scaled time decays, non-Markovian effects and also non-Gaussian noise statistics \cite{dwyer_probing_2022,LiuNV}.\\
To address these challenges, different approaches have been pursued: apart from inherently decoupling the dynamics of the system from the noisy interaction \cite{Jelezko2011,Tan2013,Barry2023},  optimization of the measurement scheme using the quantum Fisher information \cite{Cirac1997,Escher2011,Huelga2016,Jordan2017}, inference based methods \cite{ijaz2025more,Huerta2022}, quantum error correction (QEC) \cite{waldherr_quantum_2014,Dur2014,Lukin2014} and quantum error mitigation (QEM) \cite{Temme2017,Yamamoto2022,hama2023,rossini_single-qubit_2023,VanDyke2024,Liang2024,Liao2025,Wang2025} can be used to improve the experimental results.
In QEM one improves the quality of quantum experiments by leveraging information about the noise in the system at the cost of additional resources such as measurement shot numbers. The most complete information about the noise can be acquired by performing a tomography before the experiment of interest to obtain the quantum channel describing the dissipative evolution of the system. However, how to optimally use this information for QEM directly on the device to improve the measurement outcomes of quantum sensing -- especially in the presence of complex noise -- is not trivial.  \\
\begin{figure*}[t]
\centering
\includegraphics[width=0.9\linewidth]{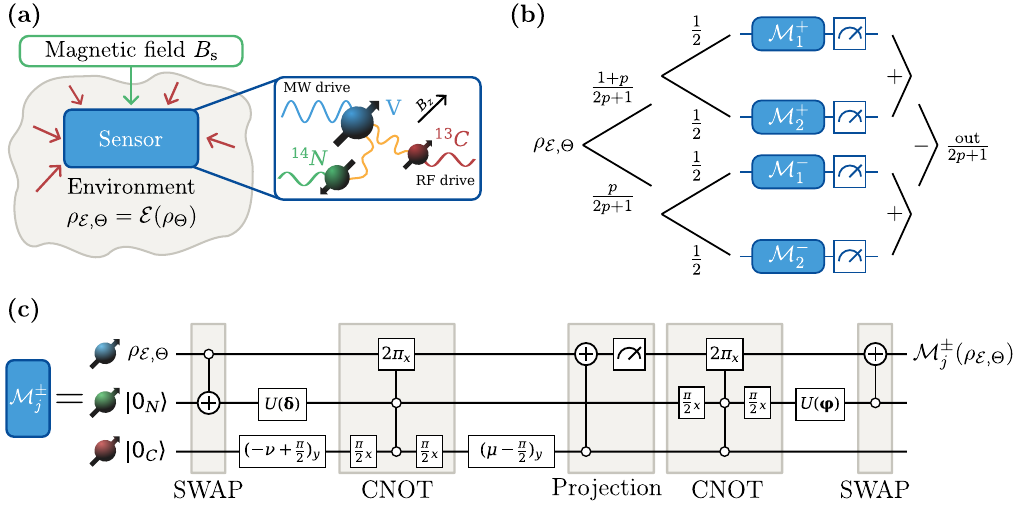}
\caption{\label{fig:circuit}
\textbf{(a)} 
Environmental noise affects the sensitivity of a sensor measuring an external magnetic field. 
\textbf{(b)} 
General scheme to implement the error mitigation map. Measurement shots are divided into the mitigation circuits according to the given probability distribution. After a projective measurement the results are collected and yield the mitigated result. 
\textbf{(c)} 
Implementation of $\mathcal{M}^\pm_j$ circuits on an NV center supported by a $^{13}$C atom. The noisy quantum information is swapped from the NV center electron spin to the $^{14}$N nuclear spin, allowing the electron spin to be reinitialized.  The information of the noise to be mitigated only enters via the angles $\boldsymbol{\delta}$, $\boldsymbol{\varphi}$, $\nu$ and $\mu$. CNOT gates between two nuclear spins are implemented as proposed in~\cite{waldherr_quantum_2014}. During the circuit the $^{13}$C nuclear spin is projected into one of its two eigenstates via a single measurement of the NV center electron spin.}
\end{figure*}
In this Letter, we address this problem by proposing a novel  model-free QEM technique for quantum sensing capable of inverting the effects of general single-qubit noise sources that are described via a CPTP (Completely Positive and Trace Preserving) channel $\mathcal{E}$. This includes multi-timescale non-Markovian processes without assumptions of a specific type of noise model.
We show how the information of the map $\mathcal{E}$ - obtained via  noise tomography - can be leveraged to derive a mitigation map for quantifying and achieving the best theoretical sensitivity. This technique is based on the fundamental concept that any mitigation process can be expressed by a combination of two engineered CPTP maps. We show how to realize these processes on a quantum device by averaging over (at most) four different quantum circuits and  indicate how the scheme can be optimized and simplified for certain types of noise.
In addition, we propose how the circuits can be efficiently implemented on single NV-center sensors by utilizing surrounding nuclear spins.
We validate the method by numerically testing it against different noise sources in DC and AC magnetometry, thus demonstrating its efficiency as a general method for error mitigation in quantum sensing applications. 
\paragraph{Sensing with a single NV center--}
In standard NV sensing applications, Ramsey interferometry is performed to measure the intensity of an unknown magnetic field. Here, the NV is initialized in a superposition of its eigenstates with energy difference $\hbar \omega_0$, enabling an accumulation of a phase difference $\Theta$ between the superposed states which is sensible to the presence of an external magnetic field $B_s$.
$\Theta$ is subsequently mapped to a population difference between the two basis states, allowing to determine the magnetic field $B_s$ with high accuracy (for details see Appendix \textcolor{red}{A}). \\
In practice, the NV center dynamics is affected by the surrounding environment during the sensing process, which in QEM is typically considered stable over the time-span of several experiments  \footnote{For NV-centers this stability comes from the localized spin environment and from the phonon bath which can be considered in equilibrium at a specific temperature. For shallow NV-centers the de-localized surface spins are sufficiently described by ensemble averaging over different spin configurations on the surface, making this also constant over time \cite{dwyer_probing_2022}. }. 
Generally, the action of the environment is determined by its spectral density around the frequency of the NV center and can be described by a completely positive and trace-preserving (CPTP) map $\mathcal{E}$ if the bare NV-center is initialized in a pure state before the Ramsey protocol. 
If the Larmor frequency of the unknown magnetic field is small compared to the frequency $\omega_0$ of the NV center (i.e. $\gamma_e B_s\ll  \omega_0 $ with $\gamma_e$ the gyromagnetic ratio of the electron spin), the spectral density of the environment can be assumed to be constant within the interval $[\omega_0-\gamma_eB_s,\omega_0+\gamma_e B_s] $. This makes the action of the environment independent of the sensing field and leads to a final noisy state $\rho_{\mathcal{E},\Theta}= \mathcal{E}(\rho_{\Theta})$, where $\rho_\Theta= U_\Theta \rho_0U_\Theta^\dagger$ is the expected outcome of the Ramsey protocol after accumulation of the phase $\Theta$ in the absence of noise. 
Here and in the following, we omit the explicit dependence of $\Theta$ and $\mathcal{E}$ on the sensing time $\tau$ in the notation.
The signal measured after the sensing protocol is $S_{\mathcal{E}, \Theta} = \rm{Tr}(\rho_{\mathcal{E},\Theta}\sigma_z)$, where $\sigma_z$ is the observable encoding the population difference. \\
\begin{figure*}[ht]
\centering
\includegraphics[width=1.0\linewidth]{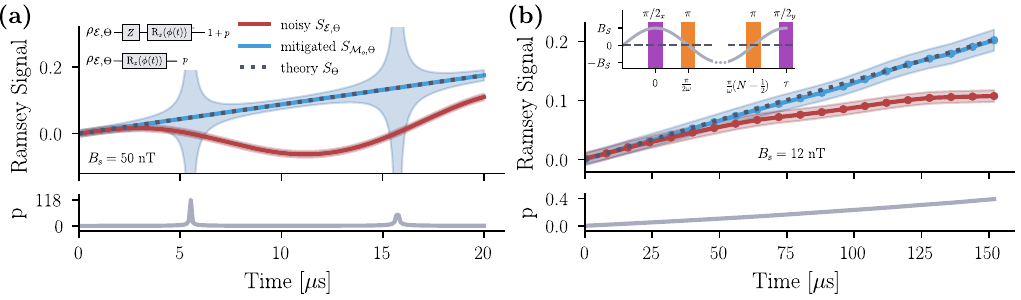}
\caption{\label{fig:main_plot}
Ramsey signal as a function of different sensing times $\tau$, without applying the mitigation circuit $S_{\mathcal{E}, \Theta}$ (red), the ideal signal $S_\Theta$ in absence of noise  (black dotted)  and  the signal calculated as a result of the mitigation $S_{\mathcal{M},\Theta}$ (blue). Shaded regions indicate the standard deviation of the result of the mitigation for $N = 10,000$ shots. The lower plots show the corresponding value of $p$ for each sensing time $\tau$ which, consistently, is closely correlated to the standard deviation. In both cases a mitigation map $\mathcal{M}_O$ which minimizes p and therefore the standard deviation is chosen. \textbf{(a)} Simulation results for DC magnetometry in presence of a surface spin bath. For this type of noise the mitigation circuits simplify as shown in the inset. \textbf{(b)} Sensing of an AC magnetic field with an NV center within a phononic environment.
}
\end{figure*}
\paragraph{Noise mitigation--}
Determining a phase $\Theta$ from a measurement in the presence of noise is, in general, a non-trivial task.
Here, we achieve this by applying a mitigation map $\mathcal{M}$ after the sensing protocol, allowing us to restore the noiseless signal $S_\Theta = \rm{Tr}(\rho_\Theta\sigma_z)$ by leveraging the information gained from the noise pre-characterization.
For this, we define a set of maps 
\begin{equation}
    M\!\!:=\!\!\{\mathcal{M}~\!|\!~\rm{Tr}(\mathcal{M}(\rho_{\mathcal{E},\Theta})\sigma_z) \!=\! \rm{Tr}(\mathcal{E}^{\text{-}1}(\rho_{\mathcal{E},\Theta})\sigma_z)\},\label{Eq.one}
\end{equation}
such that  the  signal after mitigation is given by $S_{\mathcal{M}, \Theta} = \rm{Tr}(\mathcal{M}(\rho_{\mathcal{E},\Theta})\sigma_z)$. Please note that this is equivalent to $\rm{Tr}(\mathcal{M}(\sigma_i)\sigma_z ) = \rm{Tr}(\mathcal{E}^{\text{-}1}(\sigma_i)\sigma_z )$ for any $\sigma_i$.
While $S_{\mathcal{M}, \Theta}$ does not depend on the choice of $\mathcal{M} \in M$ by construction, the standard deviation of the signal does (see below), yielding an improved sensitivity for certain choices of $\mathcal{M}$.
$\mathcal{M}$ is generally not a completely positive map, and therefore cannot be directly implemented on a circuit by standard methods. Nonetheless, it can be decomposed into a weighted difference of two CPTP maps via
$\mathcal{M}=(1+p)\mathcal{M}^+-p\mathcal{M}^-$, where $p$ is a positive weighting factor.   
Here, we express $\mathcal{M}$ by  dividing each CPTP map into a  weighted sum of two \textit{extremal} CPTP maps  $\mathcal{M}_{1,2}^+$ and $\mathcal{M}_{1,2}^-$  (see Fig.~\ref{fig:circuit}) \cite{rossini_single-qubit_2023}. 
The sensing signal after mitigation is then given by 
\begin{align}
S_{\mathcal{M}, \Theta} \!= &\frac{(1+p)}{2}(S^+_{\Theta 1}+S^+_{\Theta 2} ) \! -\!\frac{p}{2}( S^-_{\Theta 1}+S^-_{\Theta 2})  ,
\label{eq:mitigated_signal}
\end{align}
where  $S^\pm_{\Theta j} = \rm{Tr}\{\mathcal{M}_j^\pm(\rho_{\mathcal{E},\Theta})\sigma_z)\} $  denotes the expectation value of the respective mitigation circuit.
Each extremal map $\mathcal{M}_j^\pm$ can then be applied to $\rho_{\mathcal{E},\Theta}$ by 
a simple quantum circuit using just one ancillary qubit \cite{wang_solovay-kitaev_2013}.
The derivation of the above formula and the method for tuning the circuit parameters accordingly is summarized in Appendix \textcolor{red}{B}. Moreover, in Appendix \textcolor{red}{C} we demonstrate that some scenarios allow to implement some of the extremal CP maps without the need of an ancillary qubit. There, we furthermore analytically characterize the decomposition for pure dephasing, relaxation and thermalization channels, discussing the optimal resources required for their mitigation. \\
To implement the mitigation scheme on an NV center based quantum register, we leverage
the presence of the $^{14}N$ and $^{13}C$ nuclear spins (Fig.~\ref{fig:circuit}). Swapping the NV spin-state information onto the $^{14}N$ nuclear spin after the sensing protocol, we allow for repetitive quantum non-demolition measurements using ancilla-assisted repetitive readout techniques \cite{neumann_single-shot_2010, dreau_single-shot_2013, lovchinsky_nuclear_2016} while using the $^{13}C$ nuclear spin as ancilla qubit.
Because the combination of the unitary gates and the projection operation is executable on the order of a few hundreds of microseconds, the single-shot readout of the nuclear spin at the end of the mitigation circuit is the time-limiting factor for the protocol, requiring around $5~\mathrm{ms}$ ~\cite{neumann_single-shot_2010}.  
While in theory the mitigation circuits allow us to perfectly restore the ideal signal, in practice the finite amount of measurement shots $N$ limits the precision with which the noiseless signal can be inferred. Hence, in this context a detailed discussion of the achievable sensitivities is necessary. \\
\paragraph{Sensitivity--}
Following \cite{barry_sensitivity_2020}, we define the sensitivity $\eta$ of a measurement as $\eta = \delta B \sqrt{N\tau}$, where $\delta B$ is the change in magnetic field for which the signal is equal to its standard deviation. For our mitigation scheme we find the sensitivity:
\begin{align}
 \eta_{\mathcal{M}} = \frac{\sqrt{N \tau}\Delta S_{{\mathcal{M}},\Theta}}{|\partial_{B_S} (\Theta) \partial_\Theta S_{{\mathcal{M}},\Theta}|},
\end{align}
where $\Delta S_{{\mathcal{M}},\Theta}$ is the standard deviation of $S_{{\mathcal{M}},\Theta}$. 
In what follows, we consider the case of sine magnetometry
\cite{barry_sensitivity_2020}. Here, the ideal signal for small magnetic fields is given by
$
   S_\Theta =  \mathrm{Tr}(\rho_\Theta \sigma_z) = \sin(\Theta) =\Theta + \mathcal{O}(\Theta^3),
$
leading to $\partial_\Theta S_{{\mathcal{M}},\Theta}\approx 1$ for $\Theta \ll 1$. 
$\Delta S_{{\mathcal{M}},\Theta}$ can therefore be derived from Eq.~(\ref{eq:mitigated_signal}) yielding
\begin{align}
\Delta S_{\mathcal{M},\Theta}&=\sqrt{\frac{(2p+1)}{2N}}\left[p(2-(S^-_{\Theta1})^2-(S^-_{\Theta 2})^2) \right.  \nonumber\\
&\left.+ (1+p)(2-(S^+_{\Theta1})^2-(S^+_{\Theta 2})^2) \right]^{1/2},
\end{align}
where we distribute the number of shots $N$ to the four circuits according to the value $p$ to optimize the sensitivity (see Fig.~\ref{fig:circuit}) \footnote{The number of shots for the \textit{minus channels} is $\frac{p}{2(2p+1)} N$.}.  We can now use this to define the upper bound 
\begin{align}
    \eta_{\mathcal{M}} \leq  \frac{\sqrt{\tau}(2p +1)}{|\partial_{B_S}(\Theta)|} \label{eq:bound}
\end{align}
which is directly proportional to the weighting factor $p$. 
For a constant magnetic field, we find $\Theta = \gamma_e B_S \tau$, leading to
$
    \eta_{\mathcal{M}} \leq  {(2p +1)}/(\gamma_e \sqrt{\tau}). \label{eq:bound2}
$  
Optimizing the choice of the undetermined parameters of $\mathcal{M}$ to obtain the smallest value of $p$ results in an optimal mitigation map $\mathcal{M}_O$ that minimizes the sensitivity $\eta_{\mathcal{M}}$.
The best sensing time to achieve optimal sensitivity can now be directly inferred from the noise pre-characterization of $\mathcal{E}$ at different times $\tau$ using the relation given above. In relatively simple cases, such as purely dephasing effects, the value of $p$ is directly related to $e^{\tau/T_2^\star}$ (see Appendix \textcolor{red}{C1}). In contrast, the introduced method can be interpreted as a general scheme for the determination of an optimal $p$ from the noise at time $\tau$ without relying on any underlying noise model.
\\
In the following, we demonstrate the proposed error mitigation scheme for two different sensing scenarios, characterized by two different dominant sources of noise. 
Thereby, we simulate the full dynamics of a NV-center performing the Ramsey protocol for DC and AC magnetometry in a spin- and a phonon-bath, respectively. \\
\paragraph{Simulations for DC magnetometry--}
In DC magnetometry with shallow NV centers, unpaired electron spins on the diamond surface are a dominant source of noise \cite{dwyer_probing_2022}.
We describe random spin hopping on the surface via configurational averaging and assume the presence of an additional, spatially trapped electron spin that is coherently coupled to the NV center.
Using the generalized correlated cluster expansion method (gCCE), we take the microscopic structure of the 2D surface spin bath into account \cite{LiuReview,LiuNV,Galli2021,maile_performance_2024,Hahn2024}, ensuring convergence of the surface interaction radius and the number of spin bath configurations, as detailed in the Appendix \textcolor{red}{D1}. 
In this setting, the surface spins only interact with the $z-$component of the NV center electron spin, resulting in pure dephasing NV dynamics. This allows for a simplification of our general error mitigation protocol, described in Appendix \textcolor{red}{C1}, which only requires two single-qubit circuits and does not require ancillas or the SWAP operation on the nitrogen nuclear spin, as shown in the inset of Fig.~\ref{fig:main_plot}a. Notably, this way it is also possible to mitigate the noise occurring in spin-ensemble measurement devices.
We compare the measured Ramsey signal $S_{\mathcal{E},\Theta}$ without mitigation to the signal obtained after applying our protocol in Fig.~\ref{fig:main_plot}a, for a sensing field of $B_{\rm s} = 50~\mathrm{nT}$. Here, $S_\Theta=\gamma_eB_s\tau$ and the blue shaded region indicates $\Delta S_{\mathcal{M},\Theta}$ for $N = 10,000$ shots.
In this example, additionally to dephasing stemming from the delocalized spin bath, the fixed electron spin on the surface yields coherent oscillations, effectively making the impact of the environment not invertible at specific points in time. These points are signaled by the peaks in the value $p$ plotted below the Ramsey signal in Fig.~\ref{fig:main_plot}a. Interestingly, we find that the optimal $\tau$ achieving highest sensitivity is given by a value beyond the peaks. \\
To validate the performance of our mitigation method, in  Fig.~\ref{fig:sensitivities} we compare the sensitivity of the signal after the mitigation process to the sensitivity reached
in Noise Aware Quantum Sensing (NAQS), where it is assumed that the influence of the noise on the signal is perfectly known and the noise can therefore be inverted exactly (see also Appendix \textcolor{red}{E}).
We show in Fig.~\ref{fig:sensitivities}a that in the case of dephasing noise, we manage to reach the sensitivity of NAQS without requiring previous knowledge of the underlying noise model. Moreover, although this setting does not always allow us to improve the sensing signal by noise inversion with a finite number of shots, we can exactly retrieve the best possible sensitivity for the optimal sensing times. Because the noise stemming from the surface spin bath only acts in the measurement direction $\sigma_z$, we find $\eta_{\mathcal{M}_O}=\eta_{\mathcal{E}^{\textbf{-}1}}$, where $\mathcal{M}_O$ is the optimized mitigation map leading to the smallest sensitivity $\eta_{\mathcal{M}_O}$.\\
\begin{figure}[t]
\centering
\includegraphics[width=1.\linewidth]{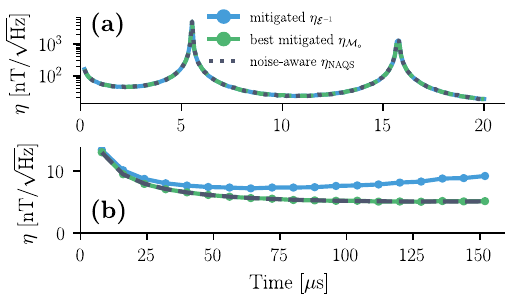}
\caption{\label{fig:sensitivities} Sensitivity as a function of sensing time $\tau$. The blue line corresponds to the sensitivity $\eta_{\mathcal{E}^{-1}}$ obtained by the mitigation technique implementing the inverted noise map $\mathcal{E}^{-1}$.
For comparison we show the sensitivity $\eta_\text{NAQS}$ reached in Noise Aware Quantum Sensing (NAQS) assuming the noise on the signal can be perfectly inverted. The green line corresponds to the sensitivity that can be reached when choosing an optimized mitigation map $\mathcal{M}_O$. \textbf{(a)} Sensitivity for DC magnetometry in presence of a surface spin bath corresponding to the signal shown in  Fig.~\ref{fig:main_plot}a (here $\eta_{\mathcal{E}^{-1}} = \eta_{\mathcal{M}_O}$). \textbf{(b)} Sensitivity for AC magnetometry in presence of a phonon bath corresponding to the data shown in Fig.~\ref{fig:main_plot}b.} 
\end{figure}
\paragraph{Simulations for AC magnetometry--}
While in DC sensing the coherence time of the NV center is limited by the interaction with the spin bath, when sensing an AC magnetic field dynamical decoupling pulses can be used to extend coherence time of the system. In this setting, the longitudinal coupling of the spin to a phononic bath becomes the prominent source of errors \cite{Norambuena2018}. 
We therefore simulate the effect of a generally non-Markovian phononic bath on a NV center while sensing an AC magnetic field and mitigate its influence on the system using. The simulations are performed using the HEOM solver provided by the QuTiP package \cite{qutip5,Lambert2023} -- see also Appendix \textcolor{red}{D2}. In this case the phase is defined as $\Theta = \int_0^\tau dt |B_s(t)| \gamma_e$. However, if we can assume several oscillations of $B_s$ during the sensing process or if the measurement is only performed after full oscillations of $|B_s(t)|$, we get $\Theta = \gamma_e \overline{B_s}\tau$, where $\overline{B_s}$ is the time average of the sensing field.
In Fig.~\ref{fig:main_plot}b we show the simulated Ramsey signal for an oscillating sensing field with amplitude 
$12~\rm{nT}$ and frequency 
$\omega_{s}/2 \pi = 62.5~\rm{kHz}$. We show that also in this scenario the mitigation  recovers the ideal signal, at the price of an increased  $\Delta S_{\mathcal{M},\Theta}$,  indicated by the blue shaded region.\\
The decomposition of the $\mathcal{E}$ map shows that it is necessary to use an ancilla qubit to achieve the best mitigation on the measurements in this scenario. A thorough explanation of the resources (ancillas) needed for the respective mitigation circuits is provided in the Appendix \textcolor{red}{C2}.
Similarly to before, we compare the simulated sensitivity to the sensitivity of NAQS in Fig.~\ref{fig:sensitivities}b. The mitigated sensitivity for $\mathcal{M} =\mathcal{E}^{-1}$ is slightly increased compared to NAQS, due to the fact that $\Delta S_{\mathcal{M},\Theta}$ is not optimally converged as $p$ is chosen to invert the full noise map. We show that this can be circumvented by choosing a different mitigation map $\mathcal{M}_O$ only inverting noise in the measurement direction, thus optimizing $p$ and the sensitivity. Thus, we obtain $\eta_{\mathcal{M}_O}< \eta_{\mathcal{E}^{\text{-}1}}$ which also perfectly overlaps with $\eta_\text{NAQS}$ as displayed in  Fig.~\ref{fig:sensitivities}b.
\paragraph{Summary and Outlook--} We propose a novel method for quantum sensing incorporating a technique to mitigate the effects of generic noise sources, and demonstrate its wide applicability in the context of quantum magnetometry measurements. We described how to experimentally implement our technique on NV sensing devices by leveraging a surrounding $^{13}C$ nuclear spin as an ancilla qubit, and verified it numerically. Furthermore, we demonstrated that the mitigation scheme can be simplified for certain types of noise, enabling direct implementation on both single and ensemble NV center sensors. 
We note that this method can be used with any single-qubit sensing device and helps overcome the sensitivity and resolution limitations of existing sensors.
Furthermore, the introduced recipe for Ramsey interferometry can be easily generalized to other sensing protocols by adjusting Eq.~(\ref{Eq.one}) to the respective measurement scheme. 

\section*{Acknowlegdments}
\begin{acknowledgments}
We thank F. Jelezko and R. Said for valuable discussions. Financial support through the BMBF within QSens (QMAT) is gratefully acknowledged.
\end{acknowledgments}

\bibliography{
  references/biblio
  }

\newpage

\appendix
\renewcommand{\thefigure}{S\arabic{figure}}
\setcounter{figure}{0}

\onecolumngrid 
\newpage

\section{Appendix}

The appendix is organized as follows: Appendix A provides supplementary details on protocols for DC and AC magnetometry, and results for different noise sources and parameters. Appendix B outlines the mathematical background of the mitigation protocol. Appendix C applies the method to common noise channels (dephasing, relaxation, and thermalization), for which the corresponding mitigation circuits can be determined analytically. Appendix D describes the simulation methods used to generate the noisy curves for the surface electron spin bath and the phononic bath shown in the main text. Finally, Appendix E contains the derivation of the sensitivity measure.

\section{A. Details about the Magnetometry Protocols}
\begin{figure}[ht]
\centering
\includegraphics[width=0.5\linewidth]{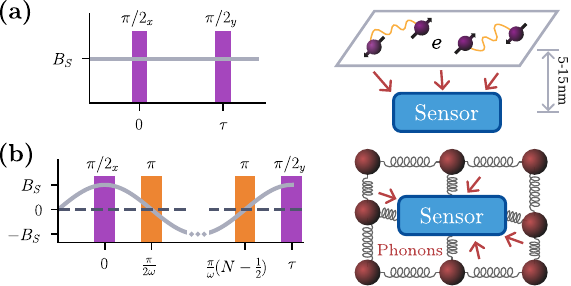}
\caption{\label{fig:A_dc_surface_ac_phonons}
\textbf{(a)} Ramsey DC magnetometry protocol. Unpaired electron spins on the diamond surface are a major source of noise for magnetometry with shallow NV centers.
\textbf{(b)} Ramsey AC magnetomatry protocol. Dynamical decoupling pulses extend the $T_2^*$ time in AC magnetic field sensing, making phononic effects relevant.
}
\end{figure}

In the simulations shown in Fig.~2 of the main text,
we used the Ramsey magnetometry protocol to sense the external magnetic field. We have considered examples of (i) DC magnetometry in the presence of electron surface spins (see Fig.~\ref{fig:A_dc_surface_ac_phonons}a) and (ii) AC magnetometry in the presence of lattice vibrations (see Fig.~\ref{fig:A_dc_surface_ac_phonons}b). In this appendix, we briefly introduce the Ramsey protocol for AC and DC magnetometry.

The Ramsey protocol for DC magnetometry consists of three steps, as shown on the left of Fig.~\ref{fig:A_dc_surface_ac_phonons}a. First, an oscillating magnetic field with amplitude $B_1$ is applied in the $x$-$y$ plane, transversely to the NV center quantization axis. In resonance with the NV center eigenfrequency, Rabi oscillations with frequency $\Omega\propto B_1$ to rotate the NV center into a superposition state by applying the transverse magnetic field for a time $\pi/(2\Omega)$. During the free precession time $\tau$ it accumulates a magnetic field dependent phase $\Theta(B_{\rm{s}}, \tau)$. In the last step, a second resonant $\pi/2$ Rabi pulse is applied with a phase shift of $\vartheta = \pi/2$ relative to the first in the $x$-$y$ plane. In this way, the accumulated phase is mapped to a population difference between the $m_s = 0$ and $m_s = \pm 1$ states that can be measured by optical readout of the NV center. Measuring the expectation value of $\sigma_z$ then leads to 

\begin{align}
    S_{\Theta} = \mathrm{Tr}(\rho_\Theta \sigma_z) = \sin(\Theta).
\end{align}

\begin{figure}[H]
\centering
\includegraphics[width=0.5\linewidth]{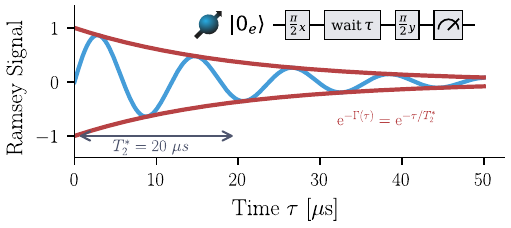}
\caption{\label{fig:ramsey}
Ramsey DC magnetic field measurement. The red curves show the decay due to dephasing effects and envelope the Ramsey oscillations. We choose the sensing magnetic field $B_{\rm s}=3~\rm\mu T$ and a dephasing time $T_2^*=20~\rm \mu s$ for visualization purposes. }
\end{figure}

Repeating this protocol for different free precession times $\tau$ yields the well-known Ramsey fringes shown in  Fig.~\ref{fig:ramsey}

For a DC magnetic field, the accumulated phase is given by $\Theta=\gamma_e B_s \tau$. When sensing small magnetic fields (and, therefore, small accumulated phases), the sine function can be linearized $S_{\Theta} = \Theta + \mathcal{O}(\Theta^3)$. In this way, the magnetic field can be determined from the slope of the Ramsey signal. In the presence of pure dephasing, the Ramsey signal is modulated by an envelope function $\exp(-\Gamma(\tau))$, e.g., $\Gamma(\tau)=2\gamma \tau = \tau/T_2^*$ for Lindblad dephasing. This yields $S_{\Theta} = \sin(\Theta) \,\rm \mathrm{exp}(-\Gamma(\tau))$ as shown in Fig.~\ref{fig:ramsey}.

In AC magnetometry, the goal is to sense the strength of a time dependent magnetic field which is given by $B(t) = B_s \cos( \omega_s t) $. Since the accumulated phase $\Theta$ is given by the time integral of the magnetic field, one needs additional decoupling pulses so that the positive and negative contributions from the cosine do not cancel each other out. Therefore, $\pi$ pulses are performed at $\tau = \frac{\pi}{\omega}(n-1/2)$ (as shown in Fig.~\ref{fig:A_dc_surface_ac_phonons}b) leading to $\Theta = \int_0^\tau \gamma_e B_s |\cos(\omega_s t)|\, \mathrm{d}t$. At the same time, the dynamical decoupling pulses cancel out contributions from static magnetic fields, effectively increasing the dephasing time of the sensor and making effects from other noise sources like surrounding phonons more prominent.

\subsection{Mitigation of Noise from Different Spin Baths}

\begin{figure}[ht]
\centering
\includegraphics[width=.9\linewidth]{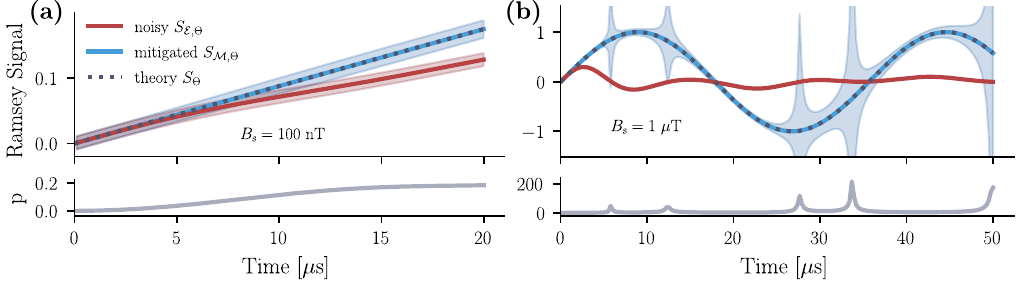}
\caption{\label{fig:main_dephasing_linear}
\textbf{(a)} Ramsey magnetometry in the linear regime $B_{\rm s}=100~\rm nT$. Similar to Fig.~2a in the main text but without a coherently coupled electron surface spin. 
\textbf{(b)} Ramsey fringes for a higher magnetic field $B_{\rm s}=1~\rm \mu T$ and surface spin density $\sigma=0.005~\rm{spins/nm}^2$. The coherently coupled electron spin fixed at position $\mathbf{r}_{\rm fix} = (2,\,0,\,10)~\mathrm{nm}$ causes the peaks in the standard deviation of the signal.  
We have used $N=10,000$ shots for both regimes.}
\end{figure}

Furthermore, we present results of the error mitigation method for dephasing noise from different electron surface spin baths: (i) without the coherently coupled electron spin, there is no peak in the standard deviation (Fig.~\ref{fig:main_dephasing_linear}a) and (ii) for higher magnetic fields and higher spin densities (Fig.~\ref{fig:main_dephasing_linear}b).

\section{B. Details about the Mitigation Protocol}

In this appendix, we describe the mathematical background of the error mitigation method presented in the main text in more detail, with a focus on possible simplifications. At its core, error mitigation involves the inversion of a noisy quantum channel $\mathcal{E}$. However, such inverse maps are generally no longer completely positive (CP) and are instead described by a general dynamical map $\mathcal{M}$. A general dynamical map is a linear, self-adjoint, and trace-preserving (TP) map, which is not necessarily CP. 

Here, we describe how to decompose and how to effectively implement general dynamical maps. The approach is based on three key steps: (1) The general dynamical map is expressed as the weighted difference of two completely positive trace-preserving (CPTP) maps \cite{rossini_single-qubit_2023}. (2) Each of these CPTP maps can, in turn, be decomposed into a convex sum of two extremal maps \cite{beth_ruskai_analysis_2002, king_minimal_2001}. (3) Extremal maps can be efficiently simulated using at most one single ancilla qubit, one CNOT gate, and single qubit rotations \cite{wang_solovay-kitaev_2013}, allowing an efficient implementation of the method. In the following, we briefly describe these steps.

\subsection*{(1) Decomposition into CPTP maps}

First, we decompose the general dynamical map $\mathcal{M}$ into a difference of CP maps $\tilde{\mathcal{M}}_\pm$. From these maps, we construct the CPTP maps $\mathcal{M}_\pm$ such that the general dynamical map $\mathcal{M}$ can be written as a weighted difference of CPTP maps. 

To achieve this, we start with the Choi matrix $C(\mathcal{M}) = \sum_{i=0}^{d^2-1}e_i\otimes\mathcal{M}(e_i)$ with $e_i$ being the matrix units and $d$ the Hilbert space dimension of the system. Since $\mathcal{M}$ is self-adjoint, this also holds for its Choi matrix, and it can be diagonalized with eigenvalues $\lambda_i$ and eigenvectors $v_i$. The Choi matrix of a CP map is positive semidefinite. Thus, by separating positive and negative eigenvalues, we can define two CP maps $\tilde{\mathcal{M}}_\pm$ by their Choi matrices (Wittstock-Paulsen decomposition)

\begin{equation}
    C(\tilde{\mathcal{M}}_\pm) = \sum_{i=0}^{d^2-1} \max(0, \pm\lambda_i)~v_iv_i^\dagger,
\end{equation}

such that $C(\mathcal{M}) = C(\tilde{\mathcal{M}}_+) - C(\tilde{\mathcal{M}}_-)$ and therefore $\mathcal{M}=\tilde{\mathcal{M}}_+ - \tilde{\mathcal{M}}_-$. Both Choi matrices are positive semi-definite by construction which implies that the underlying maps $\tilde{\mathcal{M}}_\pm$ are CP. Nevertheless, in contrast to $\mathcal{M}$ these maps are not necessarily TP anymore. 

We investigate the violation of TP by expressing $\tilde{\mathcal{M}}_\pm$ in terms of their Kraus operators. This can be achieved starting from the Choi matrix representation using $\mathrm{Tr}_{1}[C(\tilde{\mathcal{M}}_\pm)] = \sum_j K_{j,\pm}^\dagger K_{j,\pm}$. Since all linear maps acting on finite-dimensional spaces are completely bounded, this also holds for $\mathcal{M}$ and $\tilde{\mathcal{M}}_\pm$. Therefore, we can find a least upper bound for $\tilde{\mathcal{M}}_-$

\begin{equation}\label{eq:4}
    \mathrm{Tr}_{1}[C(\tilde{\mathcal{M}}_-)]  = \sum_j K_{j,-}^\dagger K_{j,-} \leq p\mathbb{I}.
\end{equation}

Since the value of $p$ determines the sampling overhead for error mitigation, this equation provides a general systematic way to improve the mitigation by optimizing $p$ as described in Eq.~(1) in the main part. Note that we have assumed finite dimensionality, i.e., the construction is not restricted to single qubits. 

If the equality holds in Eq.~\eqref{eq:4} with $p=1$, the underlying channel $\tilde{\mathcal{M}}_-$ is already TP and one can define $\mathcal{M}_\pm = \tilde{\mathcal{M}}_\pm$. Otherwise, we define the Kraus operator $D$ such that 

\begin{align}\label{eq:1}
    D^\dagger D = p \mathbb{I} - \sum_j K_{j,-}^\dagger K_{j,-}\geq 0.
\end{align}

Using the channel defined by $D$ we can define TP maps $\mathcal{M}_\pm$. We use the fact that the dynamical map $\mathcal{M}$ is TP

\begin{equation} \label{eq:2}
    \sum_j K_{j}^\dagger K_{j} = \sum_j K_{j,+}^\dagger K_{j,+} - \sum_j K_{j,-}^\dagger K_{j,-} = \mathbb{I}.
\end{equation}

Combining Eq.~\eqref{eq:1} and \eqref{eq:2} we find that the TP is fulfilled again

\begin{equation}
    \frac{ \sum_j K_{j,-}^\dagger K_{j,-} + D^\dagger D}{p} = \mathbb{I}, \qquad
    \frac{ \sum_j K_{j,+}^\dagger K_{j,+} + D^\dagger D}{1+p} = \mathbb{I}.
\end{equation}

Therefore, we can define CPTP maps $\mathcal{M}_\pm$ and have reached our goal and found a decomposition of the general dynamical map into two CPTP channels $\mathcal{M}_\pm$

\begin{equation}
    \mathcal{M}(\rho) = (1+p)\left( \frac{ \tilde{\mathcal{M}}_{+}(\rho)+D \rho D^{\dagger}}{1+p} \right)-p\left( \frac{\tilde{\mathcal{M}}_{-}(\rho)+D \rho D^{\dagger}}{p} \right) = (1+p) \mathcal{M}_+(\rho) - p \mathcal{M}_-(\rho).
\end{equation}

The maps $\mathcal{M}_\pm$ are CP because there exists a Kraus representation which automatically satisfies the CP condition. Any CPTP map can be realized by at most $d^2$ Kraus operators, which can be implemented as a unitary acting on the system and an environment of ancillary qubits (Stinespring dilation). 

For single-qubits, this means that at most four Kraus operators can be realized by at most two ancillas. However, in practical quantum sensors the coherent control of more than one ancillary qubit is often technically demanding. For this reason, we pursue a different strategy: we further decompose the desired CPTP map into extremal channels, each of which can be implemented using at most a single ancilla qubit.

\subsection*{(2) Decomposition into extremal maps}

In this step, we decompose the CPTP maps $\mathcal{M}_\pm$ into a convex sum of two extremal maps as described by Ruskai et. al \cite[Theorem 14]{beth_ruskai_analysis_2002}. A map is extremal if and only if the Choi matrix of the adjoint map is of the form 

\begin{equation} \label{eq:5}
    M = \begin{pmatrix} A & \sqrt{A} R \sqrt{B} \\  \sqrt{B} R^\dagger \sqrt{A} & B \end{pmatrix}
\end{equation}

where $R$ is a unitary (Theorem 7). Since TP implies a unital adjoint map, it holds $B=\mathbb{I}-A$. Each positive semi-definite matrix, such as the Choi matrices of the adjoint map, can be written in the form of Eq.~\eqref{eq:5} (Lemma 6). To decompose the adjoint maps into extremal maps we thus write the contraction as the equal sum of two unitaries using the singular value decomposition $R=V\Sigma W^\dagger$ (Lemma 15)

\begin{align} \label{eq:3}
    R &= V \begin{pmatrix} \cos(\theta_1) & 0 \\ 0 & \cos(\theta_2) \end{pmatrix} W^\dagger \\
    &= \frac{1}{2} V \begin{pmatrix} \exp(\rm{i} \theta_1) & 0 \\ 0 & \exp(\rm{i} \theta_2) \end{pmatrix} W^\dagger + 
    \frac{1}{2} V \begin{pmatrix} \exp(-\rm{i} \theta_1) & 0 \\ 0 & \exp(-\rm{i} \theta_2) \end{pmatrix} W^\dagger \\
    &= \frac{1}{2} U_1 + \frac{1}{2} U_2.
\end{align}

The obtained decomposition of the adjoint maps yields the decomposition of Choi matrices $C(\mathcal{M}_\pm)$ into $C(\mathcal{M}_{0,\pm})$ and $C(\mathcal{M}_{1,\pm})$ into two extremal channels $\mathcal{M}_{0,\pm}$, $\mathcal{M}_{2,\pm}$ each such that

\begin{align}
    \mathcal{M}_\pm(\rho) = \frac{1}{2} \mathcal{M}_{1,\pm}(\rho) + \frac{1}{2} \mathcal{M}_{2,\pm}(\rho)
\end{align}

We note that in practice it is useful to check if the contraction $R$ is a unitary itself such that the decomposition in Eq.~\eqref{eq:3} is not needed since the CPTP map is already extremal. This reduces the number of circuits that have to be implemented. This applies for $\mathcal{M}_\pm$ in Appendix.~C1 and for $\mathcal{M}_-$ in Appendix.~C2 such that the number of maps that must be performed by circuits reduces from four to two and three, respectively.

\subsection*{(3) Circuit realization of extremal maps}

Wang et al. \cite{wang_solovay-kitaev_2013} proposed a circuit realization of extremal maps. To determine the gates, we represent the extremal maps as Pauli transfer matrices (PTM). For CPTP maps, the PTM has the general form 

\begin{equation}
    \Lambda_P = \begin{pmatrix} 1 & \mathbf{0} \\ \mathbf{t} & \mathbf{T} \end{pmatrix}.
\end{equation}

We diagonalize the real matrix $\mathbf{T}\in\mathbb{R}^{3\times3}$ via singular value decomposition $\mathbf{T}=\mathbf{V\Sigma W}^T$, with orthogonal matrices $\mathbf{V}, \mathbf{W}$. We choose $\mathbf{V}, \mathbf{W}^T\in \rm{SO}(3)$, i.e. $\det(\mathbf{V})=\det(\mathbf{W}^T)=1$. Both matrices perform a rotation of the Bloch sphere around axis $\mathbf{n}$ and angle $\theta$ corresponding to a unitary $U=\pm\exp\left( -\rm{i} \frac{\theta}{2} \mathbf{n}\boldsymbol{\sigma} \right)$. Writing these unitaries as PTMs $\Lambda_V$ and $\Lambda_W$ we find

\begin{equation}
    \tilde{\Lambda}_P = \Lambda_V\Lambda_P\Lambda_W =  \begin{pmatrix} 1 & \mathbf{0} \\ \mathbf{0} & \mathbf{V^T} \end{pmatrix} 
    \begin{pmatrix} 1 & \mathbf{0} \\ \mathbf{t} & \mathbf{T} \end{pmatrix}
    \begin{pmatrix} 1 & \mathbf{0} \\ \mathbf{0} & \mathbf{W} \end{pmatrix} =
    \begin{pmatrix} 1 & \mathbf{0} \\ \mathbf{V^T t} & \mathbf{\Sigma} \end{pmatrix} = 
    \begin{pmatrix} 1 & \mathbf{0} \\ \mathbf{\tilde{t}} & \mathbf{\tilde{T}} \end{pmatrix}
\end{equation}

Using the trigonometric parametrization we find the particular result for extremal channels 

\begin{align}
    \mathbf{\tilde{t}} = \begin{pmatrix} 0\\0\\\sin\mu\sin\nu \end{pmatrix}, \qquad
    \mathbf{\tilde{T}} = \begin{pmatrix}
        \cos\nu & 0&0 \\ 0&\cos\mu&0 \\ 0&0&\cos\mu\cos\nu
    \end{pmatrix}
\end{align}

with $\nu\in[0,\pi)$ and $\mu\in[0,2\pi)$, such that $\sin\mu\sin\nu$ can become negative \cite{beth_ruskai_analysis_2002}. The simple form of $\mathbf{\tilde{T}}$ and $\mathbf{\tilde{t}}$ allows the realization of this map with only two Kraus operators. 

\begin{equation}\label{eq:trigono}
    K_A = \begin{pmatrix}
        \cos((\mu-\nu)/2) & 0 \\ 0 & \cos((\mu+\nu)/2)
    \end{pmatrix},
    \quad\quad
    K_B = \begin{pmatrix}
        0 & \sin((\mu+\nu)/2) \\ \sin((\mu-\nu)/2) & 0 
    \end{pmatrix}
\end{equation}

These two Kraus operators can be implemented efficiently using only one ancilla qubit, one CNOT gate, and single-qubit rotations \cite{wang_solovay-kitaev_2013}. The ancilla qubit is used to switch between the two Kraus operators. 

If only one Kraus operator is needed, one can reduce the complexity by implementing the extremal map without an ancilla. This simplification can be performed whenever the Pauli transfer matrix $\tilde{\Lambda}_P$ is unitary. This applies for $\mathcal{M}_\pm$ in Appendix.~C1, and for $\mathcal{M}_+$ in Appendix.~C2 and Appendix.~C3. 


\section{C. Application to Common Noise Channels: Analytical Solutions and Simplifications}

In this appendix, we apply the proposed error mitigation method to three different noise channels: (i) pure dephasing (phase damping), (ii) relaxation (amplitude damping), and (iii) thermalization. We explain how these noise channels can appear in the context of sensing with NV centers and give analytical solutions to construct the quantum circuits needed for mitigation. In the examples, provided we effectively implement the inverse of the noise map $\mathcal{M} = \mathcal{E}^{-1}$. We find that, in some cases, the number of mitigation circuits can be reduced and that for certain noise channels, no ancilla is required. Thus, we show that the mitigation method adapts to the type of noise.  

In the following, we use two types of transfer matrix representation for dynamical maps: the standard transfer matrix (STM) written in the computational basis of matrix units $c_i$ as $\Lambda_{c, ij} = \rm Tr\left[c_i^{\dagger} \mathcal{M}(c_j) \right]$ and the Pauli transfer matrix (PTM) written in the basis of Pauli matrices $\sigma_i$ (including the identity $\sigma_0=\mathbb{I}$) as $\Lambda_{P, ij} = \rm Tr\left[\sigma_i^{\dagger} \mathcal{M}(\sigma_j) \right]$. 

\section{C1. Mitigation of a General Pure Dephasing Channel}

Pure dephasing combined with effects from coherent coupling to trapped impurities can be described by a time-convolutionless master equation (TCL)

\begin{equation} \label{eq:TCL}  
    \frac{\mathrm d}{\mathrm d t} \rho(t) = \mathcal{L}_t[\rho(t)] = \frac{\mathrm i}{\hbar} [H(t), \rho(t)] + \frac{1}{2}\gamma(t) \left( \sigma_z\rho(t)\sigma_z - \rho(t) \right) 
\end{equation}

with time-dependent dephasing rates $\gamma(t)$ and the time-dependent superoperator $\mathcal{L}_t$. The unitary dynamics is given by the sensing magnetic field and coherent interactions with fixed bath spins acting as a time-dependent magnetic field $H(t) = \frac{\hbar}{2} (\gamma_eB_{\rm sense} + \gamma_e B_{\rm noise}(t) )\sigma_z = \frac{\hbar}{2} ( \omega_{\rm sense} + \omega_{\rm noise}(t)) \sigma_z$. 

The formal solution of the TCL master equation is given by the time-ordered exponential $\rho(t) = \mathcal{T}\exp\left(\int_0^t \mathcal{L}_{t'}\rm d t'\right) \rho_0 = \mathcal{M}_{0,t}[\rho_0]$ where the physicality requires $\mathcal{M}_{0,t}$ being CPTP for each $t$. If the time-local generator has the GKLS form with time-independent, nonnegative rates $\gamma(t)=\gamma\geq 0$, the dynamics reduces to a quantum dynamical semigroup and is CP-divisible, i.e., every intermediate map $\mathcal{M}_{s,t}$ is CPTP. Conversely, if some of the time-dependent rates become negative, the CP-divisibility condition is generally violated, which is typically taken as a signature of non-Markovian dynamics.

We use the definitions $\phi(t) \equiv \int_0^t \omega_{\rm noise}(t')~\rm d t'$ and $\Gamma(t) \equiv\int_0^t \gamma(t')~\rm d t'$ and note that for the pure dephasing channel sensing and noise can be separated $\Lambda_{c}=\Lambda_{c, \rm noise}\Lambda_{c, \rm sense}$ with

\begin{align}
    \Lambda_{c, \rm sense} = \begin{pmatrix}
        1&0&0&0\\ 0&\mathrm{e}^{\mathrm{i} \omega t}&0&0 \\ 0&0&\mathrm{e}^{- \mathrm{i} \omega t}&0 \\ 0&0&0&1
    \end{pmatrix}, \qquad
    \Lambda_{c, \rm noise} = \begin{pmatrix}
        1&0&0&0 \\0& \mathrm{e}^{\mathrm{i} \phi(t) - \Gamma(t)} &0&0 \\ 0&0& \mathrm{e}^{-\mathrm{i} \phi(t) -\Gamma(t)} &0 \\ 0&0&0&1
    \end{pmatrix}.
\end{align}

In the following, we describe how to use our error mitigation method to realize the inverse $\Lambda^{-1}_{c, \rm noise}$ and find that only two circuits without an ancilla are needed. 

\textbf{(1)} We construct the Choi matrix $C=\sum_{i=1}^4 c_i\otimes\mathcal{M}^{-1}_{\rm noise}(c_i)$ and write it as a linear combination of CPTP maps $C = \tilde{C}_+ - \tilde{C}_- = (1+p) C_+ - p C_-$. 

\begin{align}
    C &= \begin{pmatrix} 1 & 0 & 0 & \rm e^{\rm i \phi(t) + \Gamma(t)} \\ 0 & 0 & 0 & 0 \\ 0 & 0 & 0 & 0 \\ \rm e^{-\rm i \phi(t) + \Gamma(t)} & 0 & 0 & 1  \end{pmatrix} 
\end{align}

Here, it is not necessary to introduce the additional Kraus operator $D$, since $\rm{Tr}_1[\tilde{C}_-]=p\mathbb{I}$ with $p = \left( \frac{1}{2}  \rm e^{\Gamma(t)} - \frac{1}{2} \right)$ which corresponds to the equality in Eq.~\eqref{eq:4}. This implies a sampling overhead of $2p+1 = \rm e^{\Gamma(t)}$. 

\begin{align}
    C =(1+p) C_+ - p C_-&= \left( \frac{1}{2}  \mathrm e^{\Gamma(t)} + \frac{1}{2} \right) \begin{pmatrix} 1 & 0 & 0 & \rm e^{\rm i \phi(t)} \\ 0 & 0 & 0 & 0 \\ 0 & 0 & 0 & 0 \\ \rm e^{-\rm i \phi(t)} & 0 & 0 & 1  \end{pmatrix} -
    \left( \frac{1}{2}  \mathrm e^{\Gamma(t)} - \frac{1}{2} \right) \begin{pmatrix} 1 & 0 & 0 & -\rm e^{\rm i \phi(t)} \\ 0 & 0 & 0 & 0 \\ 0 & 0 & 0 & 0 \\ -\rm e^{-\rm i \phi(t)} & 0 & 0 & 1  \end{pmatrix}
\end{align}

One can quickly check if the decomposition was successful because both Choi matrices $C_\pm$ represent CPTP maps $\mathcal{M}_\pm$: TP holds since $\rm{Tr}_1[C_\pm]=\mathbb{I}$ and CP holds since $C_\pm$ are positive semi-definite (Sylvester criterion $ad-bd\geq 0$).

\textbf{(2)} We note that both Choi matrices already describe extremal maps such that for this case no further division is required. 

\begin{figure*}[ht]
\centering
\includegraphics[width=0.5\linewidth]{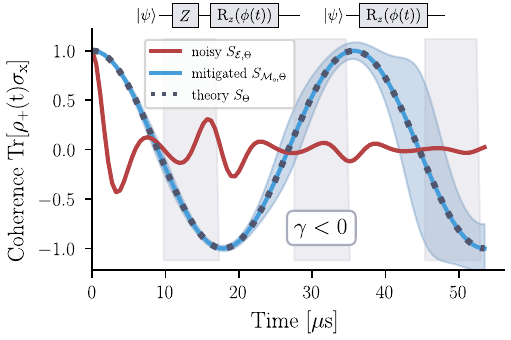}
\caption{\label{fig:dephasing}
Coherence dynamics of an NV center initially prepared in the state $\rho_0 = \ket{+}$, subject to a time-dependent dephasing rate $\gamma(t) / 2\pi = 0.3\, [\sin(2\omega_{\rm sense} t) + 0.2]~\mathrm{MHz}$ and coherent coupling to a static bath spin with strength $A_{zz} = 100~\mathrm{kHz}$. For the dephasing channel the proposed method reduces to two one-qubit circuits as shown in the inset. The target magnetic field $B_{\rm sense} = 1~\mu\mathrm{T}$ can be extracted from the mitigated curve. Blue data points represent averages over $N = 10,000$ projective measurements divided into circuits according to the probability distribution determined by the p value.}
\end{figure*}

\textbf{(3)} We see that both channels represent unitary maps for which only one Kraus operator is required such that we do not need an ancilla qubit for each of these channels. We find that the general scheme explained in Appendix~B reduces remarkably such that the CPTP maps can be realized with the Kraus operators $K_+=\rm R_z(-\phi(t))$ and $K_-=\rm Z \rm R_z(-\phi(t))$. An example is shown in Fig.~\ref{fig:dephasing}.

In NV centers dephasing is due to interactions with the surrounding spins via magnetic dipolar coupling. For shallow NV centers, typical distances from the surface electron spins are about $10~\rm nm$ such that dipolar interactions are in the order of $100~\rm kHz$ and therefore much smaller than the
resonance frequency of the NV center $\omega_0=2\pi D_{\rm gs} + \gamma_e B$ (for $|B-10^3~\rm G|\gg 1$). Therefore, the only relevant contribution comes from the $z$ couplings to the NV center which cause fluctuations in the accumulated phase. 

In the case of Lindblad pure dephasing and an effective static magnetic field (e.g., from the coherent interaction with a fixed and isolated surface spin) we get $\gamma(t)/2 = \gamma/2=1/(2T_2^*)$ and $\Gamma(t)=t/T_2^*$ as well as $B_{\rm noise}(t) =B_{\rm noise}=2\pi A_{zz}/\gamma_e $ and $\phi(t)=2\pi A_{zz}t$. 

\section{C2. Mitigation of a Relaxation Channel} 

We treat relaxation effects in a similar way as the dephasing in the previous section and describe the dynamics with a TCL master equation with relaxation rates $\gamma(t)$

\begin{equation}    
    \frac{\mathrm d}{\mathrm d t} \rho(t) = \frac{\mathrm i}{\hbar} [H(t), \rho(t)] 
    + \gamma(t) \left( \sigma_+\rho(t)\sigma_- - \left\{\sigma_+\sigma_-, \rho(t) \right\}\right). 
\end{equation}

The unitary dynamics is governed by the same time-dependent Hamiltonian as in Appendix~C1. Again we find that we can separate the maps $\Lambda_{c}=\Lambda_{c, \rm noise}\Lambda_{c, \rm sense}$ (for sensing of a magnetic field in z direction) with

\begin{align}
    \Lambda_{c, \rm noise} &= \begin{pmatrix} 1 & 0 & 0 & 1-\mathrm e^{-\Gamma(t)} \\ 0 & \mathrm e^{\rm i\phi(t) -\Gamma(t)/2} & 0 & 0 \\ 0 & 0 & \mathrm e^{-\rm i\phi(t)-\Gamma(t)/2} & 0 \\ 0 & 0 & 0 & \mathrm e^{-\Gamma(t)}  \end{pmatrix}.
\end{align}

In the following, we show how to decompose and realize the inverse $\Lambda^{-1}_{c, \rm noise}$ using the proposed error mitigation scheme. 

\textbf{(1)} The Choi matrix of the inverse map is written as the difference of CP maps $C = \tilde{C}_+ - \tilde{C}_-$.  

\begin{align}
    C &= \begin{pmatrix} 1 & 0 & 0 & \rm e^{\rm i\phi(t) +\Gamma(t)/2} \\ 0 & 0 & 0 & 0 \\ 0 & 0 & 1-\rm e^{\Gamma(t)} & 0 \\ \rm e^{-\rm i\phi(t)+\Gamma(t)/2} & 0 & 0 & \rm e^{\Gamma(t)} \end{pmatrix}  \\
    \tilde{C}_+ - \tilde{C}_- &= \begin{pmatrix} 1 & 0 & 0 & \rm e^{\rm i\phi(t)+\Gamma(t)/2} \\ 0 & 0 & 0 & 0 \\ 0 & 0 & 0 & 0 \\\rm e^{-\rm i\phi(t) +\Gamma(t)/2} & 0 & 0 &  \rm e^{\Gamma(t)}  \end{pmatrix} - 
    \begin{pmatrix} 0 & 0 & 0 & 0 \\ 0 & 0 & 0 & 0 \\ 0 & 0 & \rm e^{\Gamma(t)} - 1 & 0 \\ 0 & 0 & 0 & 0  \end{pmatrix}
\end{align}

We note that $\tilde{C}_\pm $ are not TP. Therefore, we have to define the additional Kraus operator $D$, to obtain the CPTP maps described by the Choi matrices $C_\pm$

\begin{align}
    \rm{Tr}_1[\tilde{C}_-] &= \begin{pmatrix} 0&0 \\ 0& \rm e^{\Gamma(t)} - 1\end{pmatrix} \leq (\rm e^{\Gamma(t)} - 1)\mathbb{I} = p\mathbb{I} \\
    D &= \begin{pmatrix} \sqrt{\rm e^{\Gamma(t)} - 1}&0 \\ 0& 0\end{pmatrix}
\end{align}

Using $D$ the dynamical map can be written as weighted difference of two CPTP maps as described in Appendix~B.

\begin{align}
    C = (1+p) C_+ - p C_- &= \rm e^{\Gamma(t)} 
    \begin{pmatrix} 
        1 & 0 & 0 & \rm e^{\rm i\phi(t)-\Gamma(t)/2} \\
        0 & 0 & 0 & 0 \\ 
        0 & 0 & 0 & 0 \\
        \rm e^{-\rm i\phi(t) -\Gamma(t)/2} & 0 & 0 & 1  
    \end{pmatrix} - 
    \left(\rm e^{\Gamma(t)} - 1 \right) 
    \begin{pmatrix} 
        1 & 0 & 0 & 0 \\ 
        0 & 0 & 0 & 0 \\ 
        0 & 0 & 1 & 0 \\
        0 & 0 & 0 & 0 
    \end{pmatrix}
\end{align}

In a similar way as is described in Appendix~C1 one can check if the decomposition was successful. The Choi matrices are CP since $\Gamma(t)\geq 0$ holds for physical channels even if the dephasing rates $\gamma(t)$ take negative values at certain times.

\textbf{(2)} We note that $C_-$ is already extremal. Since this does not apply for $C_+$ we further decompose this map into two extremal maps. Therefore, we define $\theta(t)=\arccos(\rm exp(-\Gamma(t)/2))$ as in Eq.~\eqref{eq:3} such that 

\begin{align}
     C_+ &= \begin{pmatrix} 1 & 0 & 0 & \rm e^{\rm i\phi(t)}\cos(\theta(t)) \\ 0 & 0 & 0 & 0 \\ 0 & 0 & 0 & 0 \\ \rm e^{-\rm i\phi(t)}\cos(\theta(t)) & 0 & 0 & 1  \end{pmatrix} \\
     \frac{1}{2}C_{1,+} + \frac{1}{2}C_{2,+} &= \frac{1}{2} \begin{pmatrix} 1 & 0 & 0 & \rm e^{i\phi(t)+\rm i \theta(t)} \\ 0 & 0 & 0 & 0 \\ 0 & 0 & 0 & 0 \\ \rm e^{-i\phi(t)-\rm i \theta(t)} & 0 & 0 & 1  \end{pmatrix} + 
     \frac{1}{2} \begin{pmatrix} 1 & 0 & 0 & \rm e^{i\phi(t)-\rm i \theta(t)} \\ 0 & 0 & 0 & 0 \\ 0 & 0 & 0 & 0 \\ \rm e^{-i\phi(t)+\rm i \theta(t)} & 0 & 0 & 1  \end{pmatrix}
\end{align}

\begin{figure*}[ht]
\centering
\includegraphics[width=0.9\linewidth]{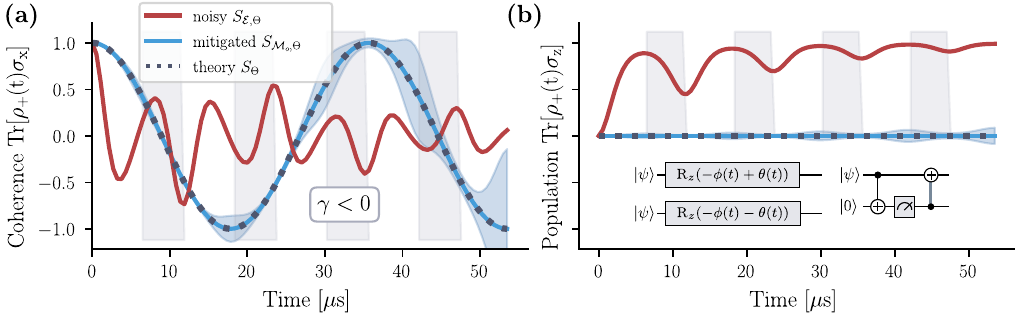}
\caption{\label{fig:relaxation}
\textbf{(a)} Coherence and \textbf{(b)} Population dynamics of an NV center exposed to relaxation noise. The system is initially prepared in the state $\rho_0 = \ket{+}$, subject to a time-dependent relaxation rate $\gamma(t) / 2\pi = 0.5\, [\sin(3\omega_{\rm sense} t) + 0.1]~\mathrm{MHz}$ and coherent coupling to a static bath spin with strength $A_{zz} = 100~\mathrm{kHz}$. For the relaxation channel the proposed method reduces to two one-qubit and one two-qubit circuits as shown in the inset. The target magnetic field $B_{\rm sense} = 1~\mu\mathrm{T}$ can be extracted from the mitigated curve in (a). Blue data points represent averages over $N=50,000$ projective measurements divided into circuits according to the probability distribution determined by the p value.}
\end{figure*}

\textbf{(3)} $C_{1,+}$ and $C_{2,+}$ belong to unitary maps and can be realized with only one Kraus operators each: $K_{1,+}=\rm R_z(-\phi(t)-\theta(t))$ and $K_{2,+}=\rm R_z(-\phi(t)+\theta(t))$. $C_-$ can be realized by $K_{1,-,A}=(\mathbb{I}+\sigma_z)/2$ and $K_{1,-,B}=\sigma_+$. Using the trigonometric parameterization from Eq.~\eqref{eq:trigono} we find $\nu=\mu=\pi/2$, such that this channel can be  realized without rotation, only using two CNOT gates performing a SWAP gate. In total, one needs two circuits without ancilla to realize $C_+$ and one with ancilla to realize $C_-$. For NV centers, $C_-$ can be realized without an ancilla by reinitializing the NV center. An example is shown in Fig.~\ref{fig:relaxation}.

Relaxation of the NV center electron spin is mainly due to lattice vibrations in the carbon crystal and can be described by a relaxation channel. 

In the case of Lindblad relaxation and an effective static magnetic field (e.g., from the coherent interaction with a fixed and isolated surface spin) we get $\gamma(t) = \gamma=1/(2T_1)$ and $\Gamma(t)=t/T_1$ as well as $B_{\rm noise}(t) =B_{\rm noise}=2\pi A_{zz}/\gamma_e $ and $\phi(t)=2\pi A_{zz}t$. 

\section{C3. Mitigation of a Thermalization Channel}

In spin-boson models the interactions of a spin with a thermal reservoir (bosonic bath) can be described by the Lindblad master equation

\begin{align}    
    \frac{\mathrm d}{\mathrm d t} \rho(t) = \frac{\mathrm i}{\hbar} [\omega\sigma_z/2  , \rho(t)]
    &+ \gamma(N+1) \left( \sigma_+\rho(t)\sigma_- - \left\{\sigma_+\sigma_-, \rho(t) \right\}\right)\\ 
    &+ \gamma N \left( \sigma_-\rho(t)\sigma_+ - \left\{\sigma_-\sigma_+, \rho(t) \right\}\right). 
\end{align}

with the Bose-Einstein thermal excitation number $N=1/(e^{\hbar \omega / k_B T} - 1)$ at temperature $T$. Even here, we find that we can separate the maps $\Lambda_{c}=\Lambda_{c, \rm noise}\Lambda_{c, \rm sense}$ for a sensing magnetic field in the z direction. With the definitions $\gamma_1 = (N+1)\gamma$, $\gamma_2=N\gamma$ and $\Gamma=\gamma_1+\gamma_2=(2N+1)\gamma$ we find

\begin{align}
    \Lambda_{c, \rm noise} &= \begin{pmatrix} (\gamma_1+\gamma_2\rm{e}^{-\Gamma t})/\Gamma & 0 & 0 & \gamma_1(1-\rm{e}^{-\Gamma t})/\Gamma  \\ 0 & \mathrm e^{\rm i\phi(t) -\Gamma t/2} & 0 & 0 \\ 0 & 0 & \rm{e}^{-\rm i\phi(t)-\Gamma t/2} & 0 \\ \gamma_2(1-\rm{e}^{-\Gamma t})/\Gamma & 0 & 0 & (\gamma_1\rm{e}^{-\Gamma t}+\gamma_2)/\Gamma  \end{pmatrix}
\end{align}

Also in this scenario, we can use the mitigation method to decompose and realize the inverse $\Lambda^{-1}_{c, \rm noise}$. 

\textbf{(1)} The Choi matrix of the inverse map is written as the difference of CP maps $C = \tilde{C}_+ - \tilde{C}_-$.  

\begin{align}
    C &= 
    \begin{pmatrix} 
        (\gamma_1+\gamma_2\rm{e}^{\Gamma t})/\Gamma & 0 & 0 & \mathrm e^{\rm i\phi(t) +\Gamma t/2} \\ 
        0 & \gamma_2(1-\rm{e}^{\Gamma t})/\Gamma & 0 & 0 \\
        0 & 0 & \gamma_1(1-\rm{e}^{  \Gamma t})/\Gamma & 0 \\
        \mathrm e^{-\rm i\phi(t) +\Gamma t/2} & 0 & 0 & (\gamma_1 \rm{e}^{\Gamma t}+\gamma_2)/\Gamma  
    \end{pmatrix} \\ 
    \tilde{C}_+ -\tilde{C}_- &= 
    \begin{pmatrix} 
        (\gamma_1+\gamma_2\rm{e}^{\Gamma t})/\Gamma & 0 & 0 & \mathrm e^{\rm i\phi(t) +\Gamma/2}\\
        0&0&0&0 \\ 
        0&0&0&0 \\ \mathrm e^{-\rm i\phi(t) +\Gamma/2} & 0 & 0 & (\gamma_1 \rm{e}^{\Gamma t}+\gamma_2)/\Gamma 
    \end{pmatrix} -
    \begin{pmatrix} 
        0&0&0&0 \\ 
        0 & \gamma_2(\rm{e}^{\Gamma t}-1)/\Gamma & 0 & 0 \\
        0 & 0 & \gamma_1(\rm{e}^{ \Gamma t}-1)/\Gamma & 0 \\
        0&0&0&0 
    \end{pmatrix}
\end{align}

We note that $\tilde{C}_\pm $ are not TP and therefore we have to define an additional Kraus operator $D$, to obtain the CPTP maps $C_\pm$

\begin{align}
    \rm{Tr}_1[\tilde{C}_-] &= 
    \begin{pmatrix} 
        \gamma_2(\rm{e}^{\Gamma t}-1)/\Gamma &0 \\
        0& \gamma_1(\rm{e}^{\Gamma t}-1)/\Gamma
    \end{pmatrix} 
    \leq (\gamma_1(\rm{e}^{\Gamma t}-1)/\Gamma) \mathbb{I} = p\mathbb{I} \\
    D &= \begin{pmatrix} \sqrt{(\gamma_1-\gamma_2) (\rm{e}^{\Gamma t}-1)/\Gamma}&0 \\ 0& 0\end{pmatrix}.
\end{align}

This way, the choi matrix can be written as a weighted difference of two CPTP maps $C = \tilde{C}_+ - \tilde{C}_- = (1+p) C_+ - p C_-$.   

\begin{align}
    C =  (1+p) C_+ - p C_- = 
    \frac{\gamma_1 \rm{e}^{\Gamma t} + \gamma_2}{\Gamma }
    \begin{pmatrix} 
        1 & 0 & 0 & \frac{\Gamma\mathrm e^{\rm i\phi(t) +\Gamma t/2}}{ \gamma_1\rm{e}^{\Gamma t}+\gamma_2 } \\
        0&0&0&0 \\ 
        0&0&0&0 \\ 
        \frac{ \Gamma\mathrm e^{-\rm i\phi(t) +\Gamma t/2} }{ \gamma_1\rm{e}^{\Gamma t}+\gamma_2 } & 0 & 0 & 1 
    \end{pmatrix} - 
    \frac{\gamma_1 \rm{e}^{\Gamma t}- \gamma_1}{\Gamma }
    \begin{pmatrix} 
        1-\frac{\gamma_2}{\gamma_1} & 0 & 0 & 0 \\
        0 & \frac{\gamma_2}{\gamma_1} & 0 & 0 \\ 
        0 & 0 & 1 & 0 \\
        0 & 0 & 0 & 0
    \end{pmatrix}
\end{align}

Similarly, as described in Appendix~C1 and C2, one can check if the decomposition was successful.

\textbf{(2)} In this case, both CPTP channels are not yet extremal. To determine the extremal map decomposition of $C_-$ we take a look at the PTM. Following the recipe, it is useful to define $\alpha=\arccos(\sqrt{\gamma_2/\gamma_1})$ and $\beta=\arccos(-\sqrt{\gamma_2/\gamma_1})=\pi-\alpha$ such that $ \sin\alpha\sin\beta = 1-\gamma_2/\gamma_1$ and $\cos\alpha\cos\beta=-\gamma_2/\gamma_1$. In order to write this as an equal superposition of two extremal channels, we introduce $\nu_1 = \mu_2 = \alpha $ and $\mu_1 = \nu_2 = \beta$. 

\begin{align}
    \Lambda_{P,-} & =
    \begin{pmatrix}
        1&0&0&0 \\
        0&0&0&0 \\
        0&0&0&0 \\
        1-\frac{\gamma_2}{\gamma_1} &0&0& -\frac{\gamma_2}{\gamma_1}
    \end{pmatrix} =
    \begin{pmatrix}
        1&0&0&0 \\
        0&0&0&0 \\
        0&0&0&0 \\
        \sin\alpha\sin\beta &0&0& \cos\alpha\cos\beta
    \end{pmatrix} \\
    \frac{1}{2}\Lambda_{P,1-} + \frac{1}{2}\Lambda_{P,2-} 
    &= \frac{1}{2} \begin{pmatrix}
        1&0&0&0 \\
        0&\cos\nu_1&0&0 \\
        0&0&\cos\mu_1&0 \\
        \sin\nu_1\sin\mu_1 &0&0& \cos\nu_1\cos\mu_1
    \end{pmatrix} \\
    &+\frac{1}{2}
    \begin{pmatrix}
        1&0&0&0 \\
        0&\cos\nu_2&0&0 \\
        0&0&\cos\mu_2&0 \\
        \sin\nu_2\sin\mu_2 &0&0& \cos\nu_2\cos\mu_2
    \end{pmatrix}
\end{align}

We also decompose the channel $C_+$ into extremal maps. To achieve this, we define $\theta(t)=\arccos(\Gamma\rm{e}^{\Gamma t/2} / (\gamma_2+\gamma_1\rm{e}^{\Gamma t}))$ as defined in Eq.~\eqref{eq:3}

\begin{align}
    C_+ &= 
    \begin{pmatrix} 
        1 & 0 & 0 & \cos(\theta(t))\rm e^{\rm i\phi(t)} \\
        0 & 0 & 0 & 0 \\ 
        0 & 0 & 0 & 0 \\
        \cos(\theta(t))\rm e^{-\rm i\phi(t)} & 0 & 0 & 1  
    \end{pmatrix} \\
    \frac{1}{2}C_{1+} + \frac{1}{2}C_{2+} &=
    \frac{1}{2}
    \begin{pmatrix} 
        1 & 0 & 0 & \rm e^{\rm i\phi(t) + \rm i\theta(t)} \\
        0 & 0 & 0 & 0 \\ 
        0 & 0 & 0 & 0 \\
       \rm e^{-\rm i\phi(t)+ \rm i\theta(t)} & 0 & 0 & 1  
    \end{pmatrix} +
    \frac{1}{2}
   \begin{pmatrix} 
        1 & 0 & 0 & \rm e^{\rm i\phi(t) - \rm i\theta(t)} \\
        0 & 0 & 0 & 0 \\ 
        0 & 0 & 0 & 0 \\
        \rm e^{-\rm i\phi(t) - \rm i\theta(t)} & 0 & 0 & 1  
    \end{pmatrix}
\end{align}

\begin{figure*}[ht]
\centering
\includegraphics[width=0.9\linewidth]{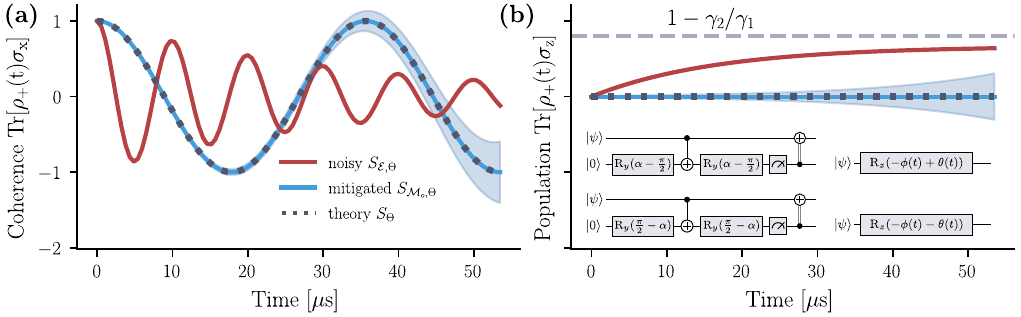}
\caption{\label{fig:thermalization}
\textbf{(a)} Coherence and \textbf{(b)} Population dynamics of an NV center exposed to thermalization noise. The system is initially prepared in the state $\rho_0 = \ket{+}$, subject to a emission rate $\gamma_1 = (N+1) \gamma$ and absorption rate $\gamma_2 = N \gamma$ and coherent coupling to a static bath spin with strength $A_{zz} = 100~\mathrm{kHz}$. For the thermalization channel all four circuits of the error mitigation method are needed, but two circuits do not require an ancilla as shown in the inset. The target magnetic field $B_{\rm sense} = 1~\mu\mathrm{T}$ can be extracted from the mitigated curve in (a). Blue data points represent averages over $N=10,000$ projective measurements divided into circuits according to the probability distribution determined by the p value. The dashed grey line indicates thermal equlibrium determined by the rate balance. }
\end{figure*}

\textbf{(3)} We notice that both extremal channels for $C_+$ are already unitary, such that an ancilla is not needed. $K_{1,+}=\rm{R}_z(-\phi(t)+\theta(t))$ and $K_{2,+}=\rm{R}_z(-\phi(t)-\theta(t))$. The extremal channels for $C_-$ are given in the trigonometric parametrization, such that they can be expressed by the Kraus operators given in Eq.~\eqref{eq:trigono}. An example is shown in Fig.~\ref{fig:thermalization}. The Kraus operators for $C_-$ are explicitly given as

\begin{align}
    K_{1,-,A}= 
    \begin{pmatrix}
        \sin\alpha &0 \\ 0&0
    \end{pmatrix}, \quad
    K_{1,-,B}= 
    \begin{pmatrix}
        0&1 \\ \cos\alpha &0
    \end{pmatrix} ,\quad
    K_{2,-,A}= 
    \begin{pmatrix}
        \sin\alpha &0 \\ 0&0
    \end{pmatrix}, \quad
    K_{2,-,A}=
    \begin{pmatrix}
        0&1 \\ -\cos\alpha &0
    \end{pmatrix} . 
\end{align}

\section{D. Details about the Simulation Method}

In this appendix, we explain how we have performed the simulations of the noisy curves in Fig.~2 in the main text.  

\section{D1. Simulation of the Surface Electron Spin Bath}

\begin{figure*}[ht]
\centering
\includegraphics[width=0.5\linewidth]{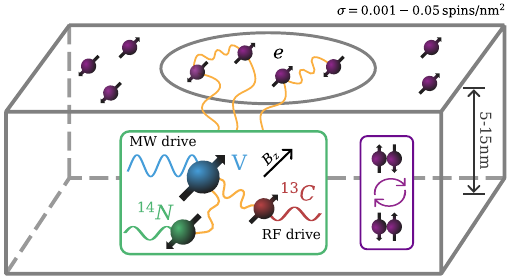}
\caption{\label{fig:visualization}
Near-surface $^{14}$NV center with the $^{13}$C nuclear spin used as ancilla in the error mitigation method. Surface electron spins are shown in purple and can undergo flip-flop interaction. An external magnetic field is applied parallel to the [111] crystal direction.}
\end{figure*}

Single NV centers placed only a few nanometers below the diamond surface are essential for high-resolution imaging. However, at the same time the NV center becomes increasingly susceptible to additional decoherence mechanisms, particularly due to unpaired electron spins at the diamond surface \cite{dwyer_probing_2022}, as illustrated in Fig.~\ref{fig:visualization}. For NV depths below $25~\mathrm{nm}$, these rapidly fluctuating surface spins dominate decoherence, outweighing the influence of other impurities such as P1 centers and C13 nuclear spins. Here, we describe the simulation methods for electron spin surface baths. 

\subsection{Convergence of the Spin Bath}

To simulate NV center decoherence caused by a surface electron spin bath with spin area density $\sigma$, we choose a cutoff radius $r_{\rm cut}$ for spins included in the simulations. We sample the number of surface spins inside this circle from a Poissonian distribution with a mean spin number $\lambda = \langle N\rangle = \pi r_{\rm cut}^2 \sigma$. Typically the spin density ranges from $0.001$ to $0.05~\rm{spins/nm}^2$ \cite{dwyer_probing_2022}. We have chosen three densities from this range: (i) low density $\sigma=0.001~\rm{spins/nm}^2$, (ii) intermediate density $\sigma=0.005~\rm{spins/nm}^2$, (iii) high density $\sigma=0.02~\rm{spins/nm}^2$. The positions of these spins are then sampled uniformly within this area. Each of these samples is referred to as a bath configuration in the following. Following \cite{dwyer_probing_2022}, we assume that surface spin positions remain fixed during individual measurements but reconfigure between measurements, a scenario known as hopping surface spins.  

\begin{figure}[ht]
\centering
  \includegraphics[width=0.5\linewidth]{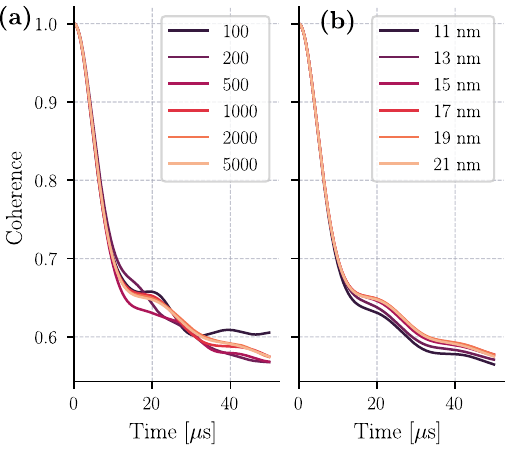}
\caption{\label{fig:converge}
\textbf{(a)}
FID for different configuration numbers. Convergence is reached at about $N=2000$ configurations.
\textbf{(b)}
FID for different cutoff radii. Convergence is reached for a cutoff radius of about $r_c = 17~\rm nm$. }
\end{figure}

\begin{figure}[ht]
\centering
  \includegraphics[width=0.5\linewidth]{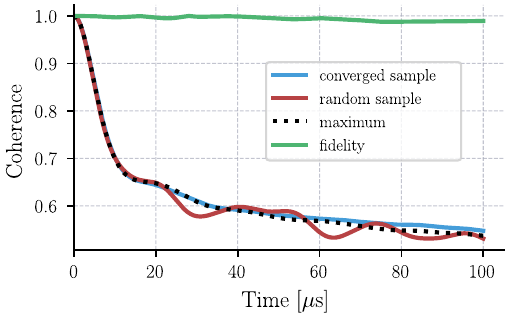}
\caption{\label{fig:converge_test}
Test of the converged parameters ($N=2000$, $r_c = 17~\rm nm$). We compare the converged parameters to the maximum we could simulate and to one random sample of less configurations and a smaller cutoff radius. The difference between maximum and the converged sample is shown as fidelity. }
\end{figure}

Both the cutoff radius and the number of bath configurations have to be converged. Therefore, we increase the values until the dynamics does not change on the timescales of the Ramsey protocol. This is shown explicitly for the intermediate spin density $\sigma=0.005~\rm{spins/nm}^2$ in Fig.~\ref{fig:converge} and \ref{fig:converge_test}.

The surface spin bath used for the simulation for Fig.~2a in the main text was generated with an intermediate spin density of $\sigma = 0.001~\mathrm{spins/nm^2}$ (low spin density). The NV center was located at a depth of $d_{\rm NV} = 10~\mathrm{nm}$, and bath spins were sampled within a cutoff radius of $r_{\rm cut} = 40~\mathrm{nm}$. We averaged over 2000 random spin bath configurations. Initial states were sampled with equal weight corresponding to the thermal distribution at room temperature. A static bias magnetic field of $B_0 = 0.05~\mathrm{T} = 500~\mathrm{G}$ was applied for the simulations.

\subsection{Spin Bath in Mean-field Approximation}

\begin{figure}[ht]
\centering
\includegraphics[width=0.9\linewidth]{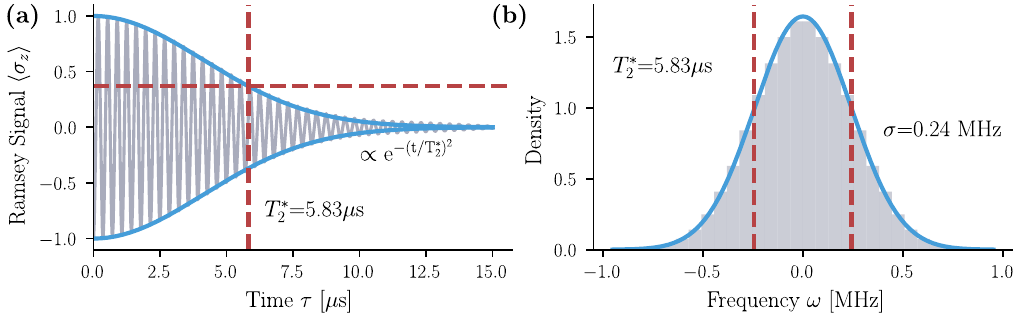}
\caption{\label{fig:ramsey_fid}  
Dephasing caused by a surface electron spin bath in mean-field approximation. The NV center is positioned $10~\mathrm{nm}$ below the surface, interacting with a bath with spin density $\sigma = 0.01~\mathrm{spins/nm}^2$ within a cutoff radius of $10~\mathrm{nm}$. The results are averaged over $10^5$ random bath configurations and all possible initial bath states. All simulations were performed under an external static magnetic field of $B_0 = 0.5~\mathrm{T}$.  
\textbf{(a)}
The sensing magnetic field $B_{\mathrm{s}} = 100~\mu\mathrm{T} = 1~\mathrm{G}$ induces rapid oscillations, while the spin bath causes the exponential decay.  
\textbf{(b)}
Histogram of frequency shifts for each bath configuration, fitted with a Gaussian distribution. The standard deviation $\sigma$ determines the dephasing time via Fourier transform, yielding $T_2^* = \sqrt{2}/\sigma$.  
}
\end{figure}

We start the analysis of the surface electron spin bath neglecting intrinsic bath dynamics, i.e. using the mean-field (MF) approximation. Due to the local magnetic field created by the surface spins, each bath configuration contributes a frequency shift (Fig.~\ref{fig:ramsey_fid}b). The Fourier transform yields a Ramsey fringe envelope function of the form due to dephasing, with the dephasing time estimated as $T_2^* = \sqrt{2}/\sigma$ (Fig.~\ref{fig:ramsey_fid}a). 

In the MF approximation, the time $\tau$ at which bath spins at radius $r$ contribute a certain phase shift can be estimated from the strength of the dipolar $A_{zz}$ interaction

\begin{align} \label{eq:radius_convergence}
    \tau = \frac{8\pi^2 (d_{\mathrm{NV}}^2 + r^2)^{5/2} }{\hbar\mu_0 \gamma_e^2 (2d_{\mathrm{NV}}^2 - r^2)} = 19.2~\frac{\mathrm{ns}}{\mathrm{nm}^3} \cdot \frac{(d_{\mathrm{NV}}^2 + r^2)^{5/2}}{(2d_{\mathrm{NV}}^2 - r^2)}.
\end{align}  

\begin{figure}[ht]
\centering
\includegraphics[width=0.5\linewidth]{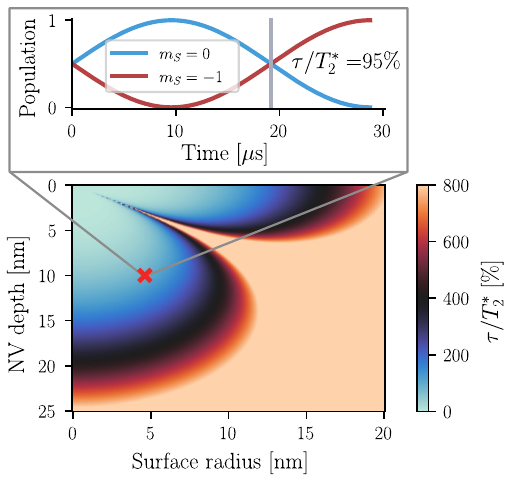}
\caption{\label{fig:radius_convergence}  
Convergence of the surface spin interaction radius as a function of NV center depth. The parameter $\tau$ represents the Ramsey free evolution time at which surface spins contribute significantly (phase $\pi$) to the NV center’s dynamics. $\tau$ is compared to the dephasing time $T_2^* = 20~\mu\mathrm{s}$.  
The inset shows Ramsey signal of an NV center located $5~\mathrm{nm}$ below the diamond surface, affected by a surface electron spin at a radius of $10~\mathrm{nm}$. From Eq.~\eqref{eq:radius_convergence}, we find $\tau \approx 19.2~\mu\mathrm{s}$, corresponding to a ratio of $\tau/T_2^* = 95\%$, marked as a red cross in the color map.
}
\end{figure}

The time $\tau$ depends on the distance between electron spin and the NV center, and therefore also on the depth of the NV center. Fig.~\ref{fig:radius_convergence} illustrates the resulting convergence of the radius.  

\subsection{Spin Bath Beyond Mean-field Approximation}

A strong external magnetic field suppresses dipolar interactions that alter spin magnetization, making the relevant dipolar interactions primarily determined by the diagonal elements of the dipolar tensor. However, surface spins with equal level splittings can undergo flip-flop interactions, mediated by the dipolar couplings $A_{xx}$ and $A_{yy}$ that are neglected in the MF approximation.

The magnetic field is assumed to be perpendicular to the diamond surface and defines the quantization axis of the surface electrons. For the sum of the flip-flop elements of the dipolar tensor, one obtains 

\begin{align}
    A_{xx} + A_{yy} \propto \frac{1}{r^3} [(3n_x^2 - 1)+(3n_y^2 - 1)] = \frac{1}{r^3}.
\end{align}

We can thus define a radius $r_{\mathrm{flip-flop}}$ within which the bath electron spins interact significantly with each other

\begin{align}
    r_{\mathrm{flip-flop}} = \left( \frac{\hbar\mu_0\gamma_e^2}{8\pi^2} \tau \right)^{1/3} \approx 3.73~\frac{\rm{nm}}{(\mu \rm{s})^{1/3}}  \cdot \tau^{1/3}.
\end{align}

\subsection{Generalized Cluster-correlation Expansion}

Exact simulations of large spin baths become computationally intractable due to the exponential scaling of Hilbert space dimensions. To account for inherent bath dynamics, we use the second-order generalized cluster-correlation expansion (gCCE) \cite{zhao_decoherence_2012, yang_longitudinal_2020, onizhuk_probing_2021, maile_performance_2024}, which captures relevant two-body interactions, including flip-flop dynamics.

\begin{figure}[ht]
\centering
  \includegraphics[width=0.5\linewidth]{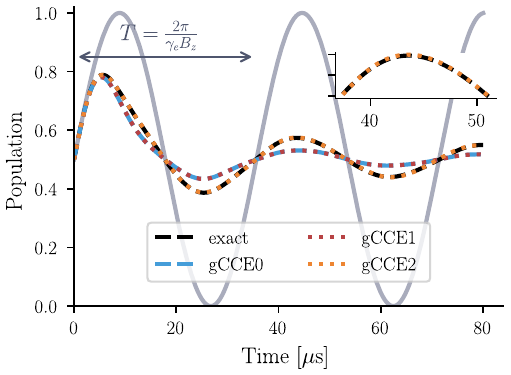}
\caption{\label{fig:gCCE_benchmark}
Ramsey fringes for a shallow NV center ($d_{\mathrm{NV}} = 10~\rm{nm}$), sensing a magnetic field $B_{\rm{sens}}=1~\mu\mathrm{T}$. The gCCE2 approximation captures two-body interactions in the spin bath, including flip-flop processes. The inset highlights residual deviations due to higher-order bath dynamics. Simulations were performed under a static magnetic field of $0.5~\mathrm{T}$, averaging over 10,000 bath configurations. }
\end{figure}

Fig.~\ref{fig:gCCE_benchmark} compares different orders of gCCE to the exact solution. While gCCE1 reproduces the mean-field result (gCCE0), gCCE2 includes two-body interactions and convergences towards the exact result.

\section{D2. Simulation of the Phononic Bath}

In the following, we will discuss the details of the simulation of a phononic bath coupled to the NV center simulating the noise in AC magnetometry. The whole system of NV center and phononic bath is described by the Hamiltonian

\begin{align}
    H = \frac{\hbar \omega_0}{2} \sigma_z + \sigma_x \sum_k \lambda_k (a_k +a_k^\dagger) + \sum_k \hbar \omega_k a_k^\dagger a_k
\end{align}

where for simplicity, we assume that the bath only couples to $\sigma_x$. Here, $\omega_0$ describes the frequency of the NV center without an additional sensing field and $\lambda_k$ the coupling to the different bath modes. Following \cite{Norambuena2018} we assume that the bath has a cubic spectral density 

\begin{align}
    J(\omega) = \alpha \frac{\omega^3}{\omega_c^2}e^{-\omega/\omega_c}
\end{align}

with dimensionless coupling constant $\alpha$ and cutoff frequency $\omega_c$ and assume that the bath is initially in a thermal state at room temperature. 
We set the NV center frequency to $\omega_0/2 \pi \approx 1.48~\rm{GHz}$, the cutoff frequency to $\omega_c =10~\omega_0$ and $\alpha = 6\cdot 10^{-9} $.

\begin{figure}[ht]
\centering
  \includegraphics[width=0.9\linewidth]{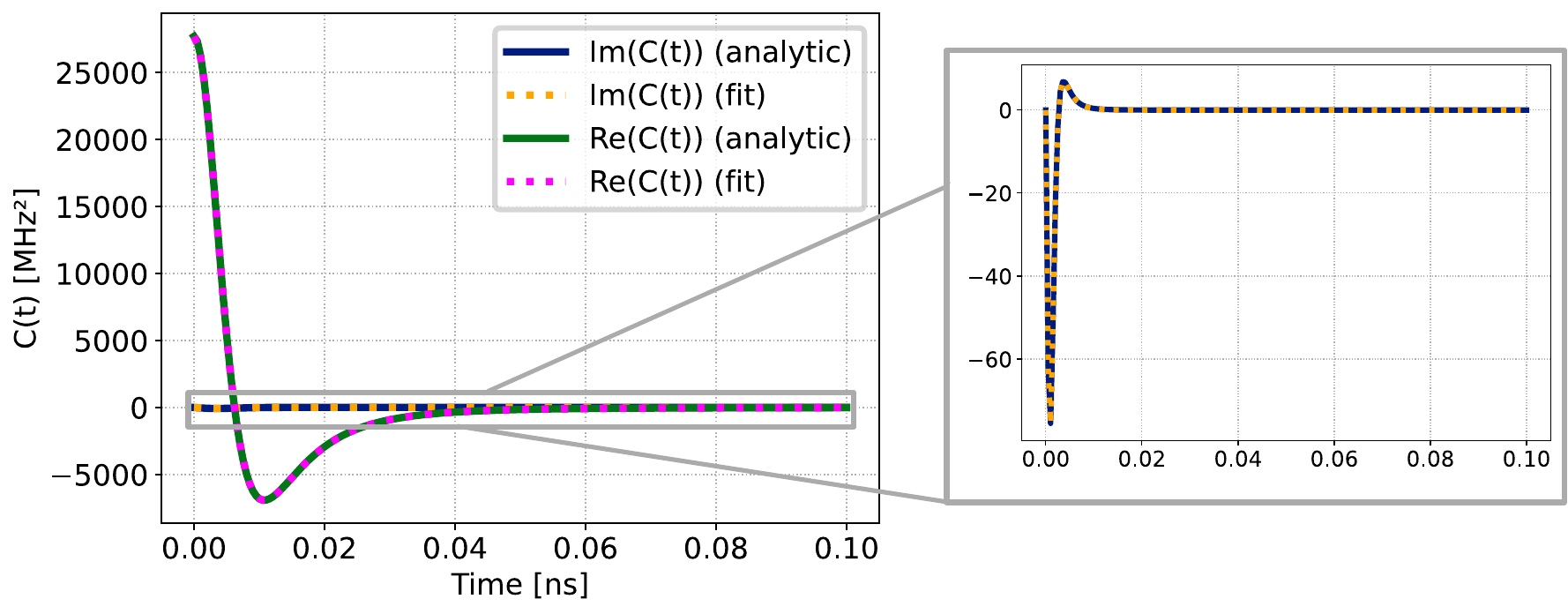}
\caption{\label{fig:fit_correlation_function}
Comparison of the fitted correlation function to the theoretical correlation function of the bath. The bath has a spectral density of $J(\omega) = \alpha \frac{\omega^3}{\omega_c^2}e^{-\omega/\omega_c}$, where  $\omega_c =10~\omega_0$, $\omega_0/2 \pi \approx 1.48~\rm{GHz}$ and $\alpha = 6\cdot 10^{-9} $. The bath is assumed to be at room temperature.}
\end{figure}

To simulate the time evolution of an NV center coupled to the phononic bath, we use the Heom solver implemented in qutip \cite{qutip5, Lambert2023}.
To do so, we fit the correlation function of the bath with a multi exponential ansatz, a comparison of the correlation function and the fitted correlation function can be seen in Fig.~\ref{fig:fit_correlation_function}.

\begin{figure}[ht]
\centering
  \includegraphics[width=0.5\linewidth]{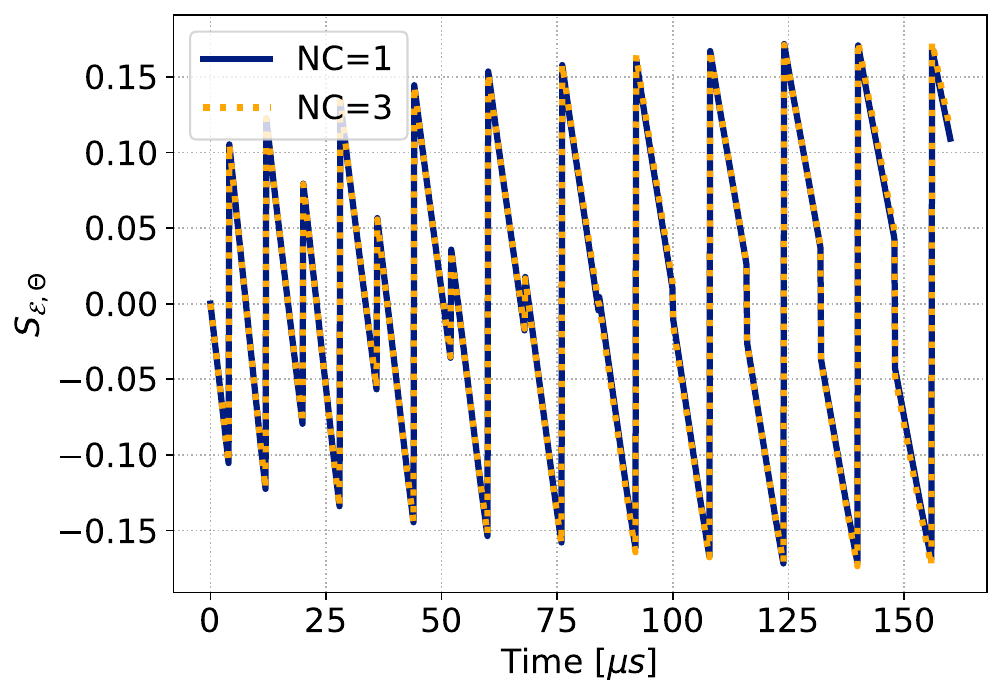}
\caption{\label{fig:fit_convergence_nc} Simulation of the signal in presence of noise and a sensing magnetic field with strength $B_S = 12 \rm{nT}$, corresponding to the data shown in Fig. 2b. Shown is a comparison of the signal simulated with a maximal depth of 1 and the signal simulated with a maximal depth of 3. As can be seen the simulations are converged. }
\end{figure}

The simulations are performed with a maximal depth of 1. To demonstrate the convergence of our result, we compare them to simulations performed with a maximum depth of 3 (see Fig. ~\ref{fig:fit_convergence_nc}).

\section{E. Sensitivity}\label{ap.E}

Lastly, we make a short summary of the derivation of the sensitivity in different sensing scenarios.

\subsection{ Noise Aware Quantum Sensing}

To analyze the sensitivity of our mitigation scheme we compare it to the case of Noise Aware Quantum Sensing, which describes the case where the influence of the noise on the signal is fully known and can therefore be inverted \cite{ijaz2025more}. In this case, the sensitivity is given by

\begin{align}
    \eta_{\rm{NAQS}} = \sqrt{N \tau}\frac{\Delta S_{{\mathcal{E}},\Theta}}{|\partial_{B_S} (\Theta) \partial_\Theta S_{\mathcal{E},\Theta}|} ,
\end{align}
with $\Delta^2S_{\mathcal{E},\Theta} = (1-(S_{\mathcal{E},\Theta})^2)/N$. Here, $N$ is the number of shots over which the signal is sampled and $\tau$ is the sensing time. 
Although in the linear regime the derivative of the ideal signal $\partial_\Theta S_\Theta$ can be approximated by 1 this is not necessarily true for the noisy signal.
However, using the linearity of the map, in the linear regime we get $S_{\mathcal{E}, \Theta} \approx \frac{1}{2} \mathrm{Tr}(\mathcal{E}(\mathbb{I})) +\frac{\Theta}{2} \mathrm{Tr}(\mathcal{E}(\sigma_z))$ so that $\partial_\Theta S_{\mathcal{E},\Theta} \approx \frac{1}{2} \mathrm{Tr}(\mathcal{E}(\sigma_z))$.
With that we obtain in the linear regime

\begin{align}
    \eta_{\rm{NAQS}} = \sqrt{\tau}\frac{\sqrt{1 -S^2_{{\mathcal{E}},\Theta}}}{|\partial_{B_S} (\Theta) \frac{1}{2} \mathrm{Tr}(\mathcal{E}(\sigma_z)\sigma_z)|} .
\end{align}

\subsection{Mitigated Sensitivity}

For our mitigation method, the sensitivity is given by

\begin{align}
 \eta_{\mathcal{M}} = \sqrt{N\tau} \frac{\Delta S_{{\mathcal{M}},\Theta}}{|\partial_{B_S} (\Theta) \partial_\Theta S_{{\mathcal{M}},\Theta}|},
\end{align}

We can now use the fact that the mitigated signal is given as a weighted sum of the signal of the four sub circuits 

\begin{align}
S_{\mathcal{M}, \Theta} \!= &\frac{(1+p)}{2}(S^+_{\Theta 1}+S^+_{\Theta 2} ) \! -\!\frac{p}{2}( S^-_{\Theta 1}+S^-_{\Theta 2})  ,
\label{eq:mitigated_signal2}
\end{align}

as well as before for the measurement of a Pauli operator, it holds
$\Delta^2S_{\Theta,i}^\pm = (1-(S^\pm_{\Theta,i})^2)/N_i^\pm$ with $N_i^\pm$ the number of shots used for each particular realization of a circuit.
To minimize the sensitivity without exact knowledge of $S_{\Theta,i}^\pm$ but assuming $(S_{\Theta,i}^\pm)^2 \ll 1$ we distribute the shots to the 4 circuits with $N_i^- = \frac{p}{2(2p+1)}N$ and $N_i^+ = \frac{p+1}{2(2p+1)}N$.
Using that 

\begin{align}
\Delta^2 S_{\mathcal{M},\Theta} = \left(\frac{1+p}{2}\right)^2(\Delta^2 S^+_{\Theta, 1}+\Delta ^2S^+_{\Theta 2} ) \! +\!\left(\frac{p}{2}\right)^2(\Delta^2 S^-_{\Theta 1}+\Delta^2 S^-_{\Theta 2}),
\end{align}

we get

\begin{align}
\Delta S_{\mathcal{M},\Theta}&=\sqrt{\frac{(2p+1)}{2N}}\left[p(2-(S^-_{\Theta1})^2-(S^-_{\Theta 2})^2) + (1+p)(2-(S^+_{\Theta1})^2-(S^+_{\Theta 2})^2) \right]^{1/2}.
\end{align}

Since the ideal mitigated signal corresponds to the noiseless signal $S_\Theta$ we can approximate $\partial_\Theta S_{\mathcal{M},\Theta} \approx 1$ in the linear regime, which leads to the sensitivity of the mitigated signal

\begin{align}
 \eta_{\mathcal{M}} = \frac{\sqrt{\tau}\sqrt{\frac{(2p+1)}{2}}\left[p(2-(S^-_{\Theta1})^2-(S^-_{\Theta 2})^2) + (1+p)(2-(S^+_{\Theta1})^2-(S^+_{\Theta 2})^2) \right]^{1/2}}{|\partial_{B_S} (\Theta) |} \leq  \frac{\sqrt{\tau}(2p +1)}{|\partial_{B_S}(\Theta)|}
\end{align}

Note that in most cases $\Theta \propto \tau$ as in DC magnetometry $\Theta = \gamma_e B_s \tau$ so that $\partial_{B_s} \Theta \propto \tau$ and

\begin{align}
    \eta_{\mathcal{M}} \leq \frac{2p+1}{\gamma_e\sqrt{\tau}}.
\end{align}

\begin{figure}[H]
\centering
\includegraphics[width=0.5\linewidth]{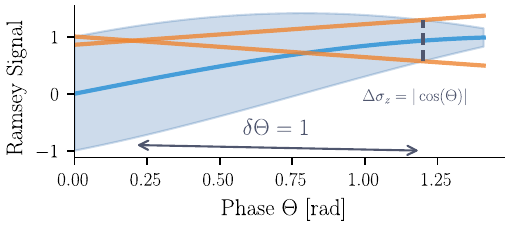}
\caption{\label{fig:ramsey2}
Visualization of the sensitivity without noise. The blue shaded region shows the standard deviation due to projection noise $\Delta S_\Theta = \cos(\Theta)$. For the largest resolvable phase $\delta\Theta=1$ the signal $\delta S_\Theta$ is equal the uncertainty $\Delta S_\Theta$. This is indicated by the crossing orange lines with $\pm$ the slope at the time of measurement. In DC magnetometry with $\delta\Theta=\gamma_e \delta B_s\tau=1$ this yields the sensitivity $\eta=\delta B_s\sqrt{\tau}= (\gamma_e\sqrt{\tau})^{-1}$}
\end{figure}


\end{document}